\documentclass[a4paper,twocolumn,11pt,accepted=2022-02-06]{quantumarticle}
\pdfoutput=1
\usepackage[utf8]{inputenc}
\usepackage[english]{babel}
\usepackage[T1]{fontenc}
\usepackage{amsmath}
\usepackage{hyperref}

\usepackage{amsfonts, amsmath, bm, bbm, graphicx, epsfig, epstopdf}
\usepackage{dcolumn}
\usepackage{textcomp}
\usepackage{soul}
\usepackage{graphicx}
\usepackage{amsfonts, amsmath, bm, bbm, graphicx, epsfig, epstopdf}
\usepackage{dcolumn}
\usepackage{textcomp}
 \usepackage[caption=false]{subfig}
\usepackage{soul}

\usepackage{tikz}
\usepackage{lipsum}

    \usepackage{braket}
\renewcommand\bra[1]{{\langle{#1}|}}
\makeatletter
\renewcommand\ket[1]{%
  \@ifnextchar\bra{\k@t{#1}\!}{\k@t{#1}}%
}
\newcommand\k@t[1]{{|{#1}\rangle}}
\makeatother

\begin{document}

\title{ Discrete-Time Quantum-Walk \& Floquet Topological Insulators  via Distance-Selective Rydberg-Interaction}

\author{Mohammadsadegh Khazali}
\affiliation{Institute for Quantum Optics and Quantum Information of the Austrian Academy of Sciences, Innsbruck, Austria}
\orcid{0000-0002-7244-3543}

\maketitle

\begin{abstract}
This article proposes the first discrete-time implementation of Rydberg quantum walk in multi-dimensional spatial  space that could ideally simulate different classes of topological insulators.
Using distance-selective exchange-interaction between Rydberg excited atoms in an atomic-array with dual lattice-constants, the new setup operates both coined and coin-less models of discrete-time quantum walk (DTQW).
Here, complicated coupling tessellations are performed by global laser that exclusively excite the site at the anti-blockade region. 
The long-range interaction provides a new feature of designing different topologically ordered periodic boundary conditions.  Limiting the Rydberg population to two excitations, coherent QW over hundreds of lattice sites and steps are achievable with the current technology.
These features would improve the performance of this quantum machine in running the quantum search algorithm over topologically ordered databases as well as diversifying the range of topological insulators that could be simulated. 
\end{abstract}

\section{Introduction}

There is a significant effort in making quantum hardwares that outperform classical counterparts in performing certain algorithms and simulating other complicated quantum systems. 
Among different approaches, implementing the quantum walk (QW) \cite{Aha93,Far98,Kem03} receives wide interest.
Unlike classical random walk, particles that perform a quantum walk can take superposition of all possible paths through their environment simultaneously, leading to faster propagation and enhanced sensitivity to initial conditions \cite{Dad18,Sum16,Pre15}. 
These properties provide an appealing basis for implementation of quantum algorithms like searching \cite{Por13, She03, Chi03,Chi04, Por17}, quantum processing \cite{Chi09,ken20,Lov10,Chi13,Sal12} and simulating the topological insulators \cite{Kit10}. 
Improving the hardware in terms of size, coherence, dimensions, controlling elements and other features like the topologically ordered boundary-conditions, would improve the performance and diverse the applications that could run over this quantum hardware.

Quantum walk has been implemented on trapped ions \cite{Sch09,Zah10}, neutral atoms \cite{Kar09,Wei11,Fuk13,Pre15} and  among other platforms \cite{Man14}. While the ion-traps are limited to a 1D array of 50 atoms, neutral atoms are pioneers in terms of multi-dimensional trapping  of a large number of identical qubits. From this perspective, trapping and  controlling  more qubits than any other platforms have already been demonstrated in one dimension (1D) \cite{Ber17,Omr19}, 2D \cite{Zha11,Pio13,Nog14,Xia15,Zei16,Lie18,Nor18,Coo18,Hol19,Sas19} and 3D \cite{Wan16,Bar18}  geometries. The current advances in exciting atoms to the Rydberg level \cite{Lev18}, provide a controllable long-range interaction for quantum technology \cite{Saf10,Ada19,Kha15,kha20, kha21, khaz2020Rydberg, khaz16,Kha21,Kha19,Kha16,Kha17,Kha18}.  
Rydberg interaction has been used in different  time-independent lattice Hamiltonian models which leads to continuous-time quantum transport \cite{Cot06,Les19,Bar15,Scho15,Ori18,Gun13,Sch15,Let18,Wus11} and simulating topological Mott insulators \cite{Dau16}. 
The continuous-time interaction in these simulators limits the range of programable problems.
Moreover, exciting all the sites to the Rydberg level would enhance de-coherence rates in the system.

This paper proposes a new approach that improves the level of control over the interaction connectivities in the lattice, leading to Rydberg DTQW in multi-dimensions.
The exchange interaction between the  quantum walker i.e. $nP$ Rydberg-excited atom and the auxiliary $nS$ Rydberg excitation at a neighboring site would delocalize the walker. The site selective excitation/de-excitation of auxiliary $nS$ state in each step, relies on the distance-dependent interaction induced level-shift and energy selection of the exciting laser.
The anisotropic dipolar interaction over a tweezer array of dimers, breaks the energy symmetry of the sites around the walker. This would insure the global laser to exclusively get in-resonance, excite $nS$, and perform quantum walk at the targeted site.  
Benefitting from the long-range Rydberg interaction, the scheme features QW implementation on topologically ordered periodic boundary conditions. Limiting the Rydberg population below two atoms  in this proposal promises scalable quantum simulation in atomic systems.  As an example, the simulation of  different classes of Floquet topological insulators with Rydberg model will be overviewed.

Topological insulators are a new class of quantum materials that insulate in the bulk but exhibit robust topologically protected current-carrying edge states.
These materials are challenging to synthesize, and limited in topological phases accessible with solid-state materials \cite{And13}.
This has motivated the search for topological phases on the systems that simulate the same principles underlying topological insulators. DTQWs have been proposed for making Floquet topological insulators. This  periodically driven system simulates an effective (Floquet) Hamiltonian that is topologically nontrivial \cite{Cay13}.  This system replicates the effective Hamiltonians from all universal classes of 1- to 3-D topological insulators \cite{Kit09,Kit10,Pan20}. Interestingly, topological properties of Floquet topological insulators could be controlled via an external periodic drive  rather than an external magnetic field.

 DTQW generates topological phases that are richer than those of time-independent lattice Hamiltonians \cite{Dau16}. Topological edge states have been realized exclusively in photonic DTQW with limited sites ($<20$) and steps ($<10$) \cite{Rec13,Xia17,Muk17}. The introduced Rydberg DTQW scheme could realize edge state over couple of hundred sites and steps in 1, 2, and 3 dimensions with the available experimental setups. The mentioned advances of the designed hardware improve the performance of quantum algorithms, e.g.  torus topological periodic boundary conditions could result in a quadratic speedup of quantum-walk-based search algorithms \cite{Por13,Por17},
  and diversify the range of topological phenomena that could be simulated. This work opens up the chance of realizing topological properties in multi-dimensional QWs as well as QWs over topologically ordered surfaces.

The article is organized as follows. In Sec.~\ref{SecScheme}, the coined and coin-less Rydberg discrete-time quantum walk schemes are presented in 1D.  Sec.~\ref{Sec_multiD_QW} extends the model to higher dimensions and discusses the approaches for imposing periodic boundary conditions or applying quantum walk on different topological surfaces.
 The Coherence of the proposed scheme under a wide range of de-coherent sources would be evaluated in Sec.~\ref{Sec_Decoherence}. 
 The scheme's performance in multi-dimensions are  then quantified in Sec.~\ref{Sec_Fidelity3D}. 
This article is supplemented by the application of the scheme in simulating multi-dimensional topological insulators.

\section{Rydberg discrete-time quantum walk}
\label{SecScheme}

\begin{figure}[h!]
\centering
\scalebox{0.42}{\includegraphics*{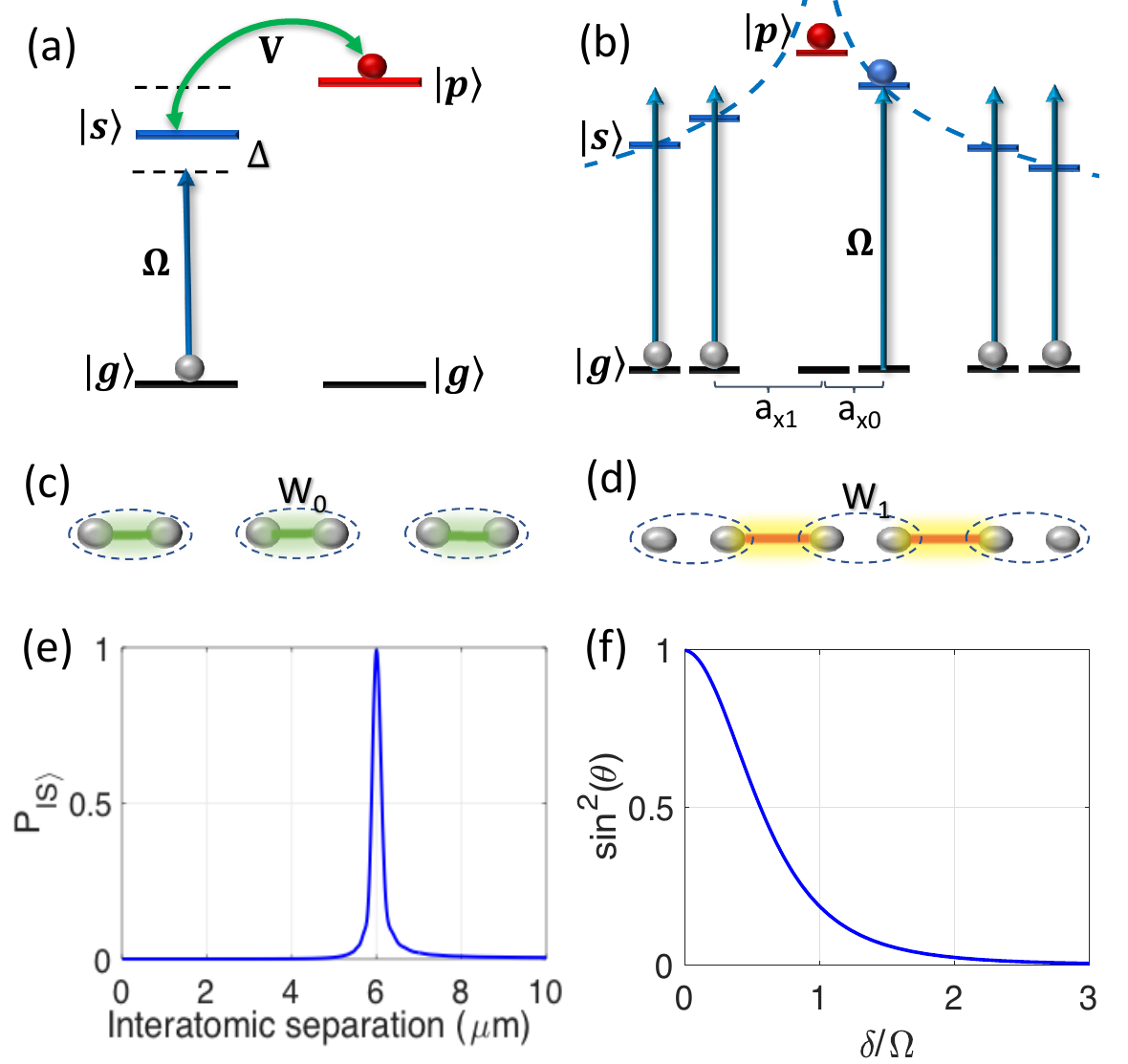}}  
\caption{ Rydberg discrete-time quantum walk (DTQW) scheme. 
(a) Level scheme: The walker is an $nP$ Rydberg excitation. QW operates by exciting a neighbouring lattice site to $nS$ Rydberg state featuring resonant exchange-interaction with the walker. 
(b) The exchange interaction $V$ forms a site-dependent level shift of $nS$ Rydberg state.
Using two lattice constants and tuning the exciting laser's frequency only a desired site get in resonance with the laser and apply quantum walk.
(c[d]) By adjusting laser's detuning to the inter- [intra-] dimer interaction, the desired coupling tessellation $W_0$ [$W_1$] would be formed.
(e) The maximum population of the auxiliary state $|100S\rangle$ over a 2$\pi$ pulse is plotted as a function of interatomic distance from the walker $|100P\rangle$. Laser detuning is set to $\Delta=-\frac{C_3}{a^3}$. Only the site at the distance $a=6\mu$m from the walker would get in resonance and hence goes under the quantum walk ($ \Omega/2\pi=2$MHz, $C_3=37$GHz.$\mu$m$^3$).
 (f)  The hopping angle $\theta$ of Eq.~\ref{Eq_W}, could be controlled by manipulating the effective detuning of the targeted site $\delta=\Delta+V$.
 }\label{Fig_Scheme}
\end{figure}

In the {\it coin-less DTQW } different coupling tessellations must be considered, in a way that each tessellation covers all vertices with non-overlapping connections and the tessellation union covers all edges \cite{Amb15,Por15,Por16,Port16,Portu17}. This model could be thought as discrete-time version of the famous Su-Schrieffer-Heeger (SSH) model \cite{su1979solitons}. Distinguishing even $\ket{m,e}=\ket{2m}$ and odd $\ket{m,o}=\ket{2m-1}$ lattice sites in the $m^{\text{th}}$ sub-lattice (dimer), the two types of QW operators depicted in Fig.~\ref{Fig_Scheme}c,d, cover the  intra-dimer $W_0$ and inter-dimer $W_1$   coupling tessellations 
\begin{equation}
\label{Eq_W}
W_0=\exp(\text{i}\theta_{0} H_{0}),  \quad \quad  W_1=\exp(\text{i}\theta_{1} H_{1}),  
\end{equation}
with $\theta$ being the hopping angle and 
\begin{eqnarray}
\label{Eq_H}
&& H_{0}=\sum\limits_{m=1}^{N/2} (\ket{m,e} \bra{m,o}+\text{h.c.})  \\ \nonumber
&& H_{1}=\sum\limits_{m=1}^{N/2} (\ket{m,e} \bra{ m+1,o}+\text{h.c.})  
\end{eqnarray}
being the coupling Hamiltonians.

The {\it physical implementation} of the proposed Rydberg discrete-time quantum walk (DTQW) is presented in Fig.~\ref{Fig_Scheme}. Here, a tweezer array of Rubidium atoms with two lattice constants is considered.
The walker is a $\ket{p}=\ket{nP_{3/2},3/2}$  Rydberg excitation while other sites are in the ground state $\ket{g}=\ket{5S_{1/2},1/2}$. The desired inter- and intra-dimer coupling labeled by $k\in \{0,1\}$, could be realized by site selective excitation of ground state atom to the auxiliary Rydberg level $\ket{s}=\ket{nS_{1/2},1/2}$  featuring exchange interaction with the walker $\ket{p}$. The site selection is controlled by adjusting of the global laser's detuning $\Delta_k$ from the auxiliary state $\ket{s}$    under the concept of Rydberg aggregate, see bellow. The effective Rydberg quantum-walk is governed under the following Hamiltonian
\begin{eqnarray}
\label{Eq_HRy}
H_{k}^{Ry}=&&\sum\limits_{i,j} V(r_{ij}) (\ket{s_i p_j} \bra{ p_i s_j}+\text{h.c.})\\ \nonumber
&&+\sum\limits_{i} \Omega/2 (\ket{s}_i \bra{ g}+\text{h.c.})+\Delta_k \ket{s}_i\bra{s},
\end{eqnarray}
where $i,j$ sums over all the sites and $V(r_{ij})$ is defined in Eq.~\ref{Eq_Vij}. Over the  2$\pi$ pulse of $\Omega$ laser, the exchange interaction $V(r_{ij})$ between the walker $\ket{p}$ and auxiliary Rydberg state $\ket{s}$ excited at the targeted site $j$, would delocalize the walker $\ket{p_ig_j}\rightarrow \cos \theta \ket{p_i g_j} + \text{i} \sin\theta \ket{g_i p_j}$.  Starting with a localized  $\ket{p}$ excitation, the first laser pulse with a specified detuning delocalizes the walker over two sites of a dimer. The laser detuning is then changed such that these two sites now interact with lattice sites on the other sub-lattices, further delocalizing the walker. Repeating this sequence would apply the coinless DTQW.

To operate the two tessellation types of Eq.~\ref{Eq_W}  under Rydberg  Hamiltonian of Eq.~\ref{Eq_HRy} with a global laser, the space-dependent nature of interaction $V(r_{ij}) $ is used over a superlattice with distinct lattice-constants inside ($a_0$) and outside ($a_1$) the dimers. By adjusting the exciting laser's detuning  from the target  $\ket{s}$ to be $\Delta_k=-\frac{C_3}{a_k^3}$, only the  lattice site at distance $a_k$ ($k\in\{0,1\}$) from the walker would get in resonance with the laser and thus undergo the quantum walk, see Fig.~\ref{Fig_Scheme}e. The single non-local quantum walker $\ket{p}$ would induce the excitation of a single nonlocal auxiliary $\ket{s}$ Rydberg state over each $2\pi$ pulse operation, justifying the absence of self interaction in Eq.~\ref{Eq_HRy}, see App.~A3. Adjusting  the laser detuning at each pulse would connect the targeted site at $a_0$ or $a_1$ distance from the walker, generating the desired  $W_k=\exp(\text{i} H^{Ry}_{k} t_k)$ with $k\in \{0,1\}$ corresponding to intra- and inter-dimer coupling tessellations in Eq.~\ref{Eq_W}.
Duration of each step $t_k$ is defined by  $\int_0^{t_k} \Omega_\text{eff} \text{d}t=2\pi $, where the effective Rabi frequency is given by $\Omega_\text{eff}=\sqrt{\Omega^2+\delta^2}$ and $\delta=\Delta_k+V(a_k)$ is the effective detuning of the targeted site at either $a_0$ or $a_1$. Fig.~\ref{Fig_Scheme}f shows how the hopping angle $\theta$ in each step can get controlled by $\frac{\delta}{\Omega}$ ratio. 
 Figure~\ref{Fig_Simulation} compares the Rydberg operation of Hadamard DTQW simulated via Eq.~\ref{Eq_HRy} with the desired Hamiltonian of Eq.~\ref{Eq_W}.

 To implement {\it coined DTQW},   the dimers would be considered as the individual units. 
 Rather than encoding the coin in electronic state, this article proposes encoding in spatial basis.
 The coin is formed by the relative population of odd and even sites in each  sub-lattice (dimer), see Fig.~\ref{Fig_Scheme}c,d.
  The coin rotation operator $R_{\theta}=\exp(\text{i} H_{0} \theta)$ is applied by population rotation in the sub-lattice basis using $H_{0}$ coupling Hamiltonian of Eq.~\ref{Eq_H}.  The desired transition operators of coined DTQW would be realized by subsequent application of intra- and inter-dimer spin transport i.e. $T=\text{e}^{\text{i}H_{1}\pi/2}R(\pi/2)=\sum\limits_{m}(\ket{m-1,e} \bra{ m,e} + \ket{m+1,o} \bra{ m,o})$.

\section{Multi-dimensional DTQW with periodic boundary conditions}
\label{Sec_multiD_QW}

\begin{figure}
\centering
\scalebox{0.4}{\includegraphics*[viewport=10 180 1000 430]{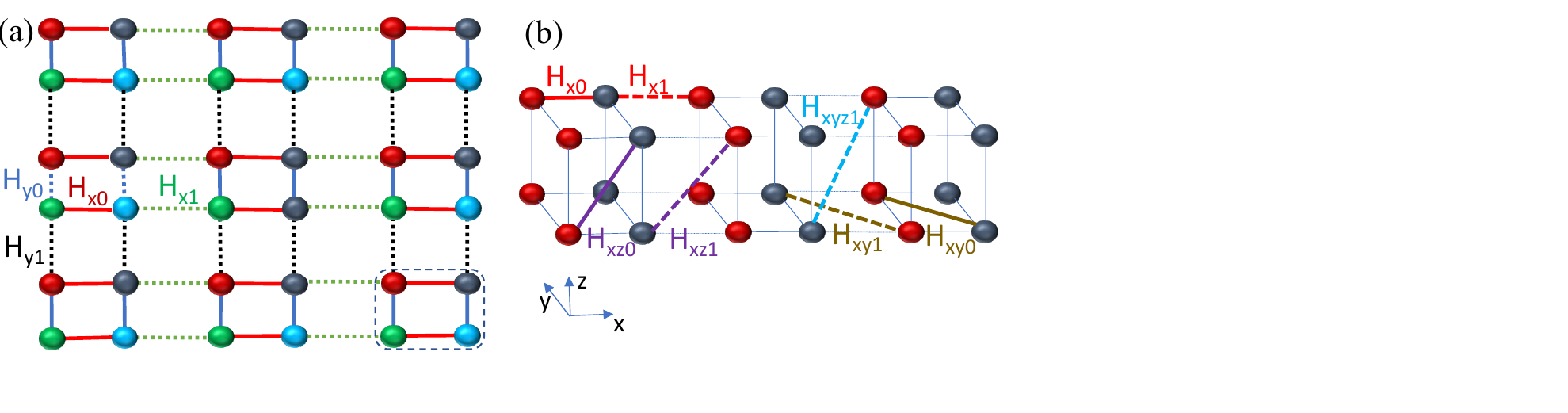}}
\caption{Multi-dimensional DTQW. (a) Kronecker  multiplication of 1D QW leads to a 2D lattice of tetramers and could trivially be extended to a 3D lattice of octamers.
(b) Extension to the 3D lattice of dimers provides a non-separable multi-dimensional Rydberg DTQW.  In (b) quantization axis would alter during the operation to be along the connections.}\label{Fig_MultiD}
\end{figure}

 The idea behind Fig.~\ref{Fig_Scheme}, is  extendable to higher-dimensions by two approaches.  Kronecker  multiplication of 1D staggered quantum walk, would make   2D and 3D lattices of  tetramers and octamers respectively. A more enriched non-separable  DTQW could be applied in a multi-dimensional lattice of dimers. 
The angular-dependency of the exchange interaction $V_{ij}$ provides a wider range of laser detuning, available for dynamic control over the exciting sites.

{\it Multi-dimensional DTQW via Kronecker  multiplication --} 
Extension to higher dimensions could be realized as the combination of  coin-less DTQWs along different directions. In two dimensions, this would result in a 2D lattice of tetramers as depicted in Fig.~\ref{Fig_MultiD}a. 
The QW is performed by concatenated application of the four sets of quantum jump operators  $W_{xl}=\exp(\text{i}\theta_{xl} H_{xl}\otimes \mathbbm{1}_{y})$  and $W_{yl}=\exp(\text{i} \theta_{yl}   \mathbbm{1}_{x} \otimes H_{yl})$ where $H_l$ ($l\in\{0,1\}$)  in each dimension is given by Eq.~\ref{Eq_H}, with distinguished odd and even sites along $x$  and $y$ dimensions, see Eq.~\ref{Eq_2DTetramers} for the expanded set of Hamiltonians. For the implementation, two lattice constants along each dimension is required to distinguish inter- and intra-cell couplings.  Extension to the 3D lattice of octamers is trivial.

{\it Multi-dimensional  Rydberg DTQW in a lattice of dimers} 
 provides an enriched non-separable Hamiltonian. The lattice structure in this model consists of unique lattice constant along $y$ and $z$ dimensions while containing two inter- and intra-cell lattice constants along the $x$ dimension. 
 Fig.~\ref{Fig_MultiD}b shows the connectivity graphs over the 3D lattice with the coupling Hamiltonians presented in Eq.~\ref{Eq_H3d}. These coupling Hamiltonians could be used for coinless DTQW operators $W_l=\text{e}^{\text{i}H_l\theta_l}$ with $l= \{x_i,xy_i,xz_i,xyz_i\}$ with $i\in\{0,1\}$ as explained in Sec.~\ref{SecScheme}.
To realize this set of couplings, the Rydberg quantization axis must be changed to be along the exchange orientation.  
The quantization axis is defined by the orientation of polarized lasers.
Fine tuning of presented connectivities and operation fidelities in multi-dimensions are discussed in Sec.~\ref{Sec_Fidelity3D}.

{\it Multi-dimensional coined DTQW  in a lattice of dimers --} 
The proposed system of Fig.~\ref{Fig_MultiD}b, can be used for the implementation of multi-dimensional coined DTQW, where the coin is formed by the relative population of odd and even sites in each sub-lattice. 
The coin rotation $R_{\theta}=\exp(\text{i} H_{x0} \theta)=\cos(\theta) \mathbbm{1}_{\{e,o\}}+\text{i} \sin(\theta) (\ket{e} \bra{ o}+\ket{o} \bra{ e})$ 
is applied by the intra-dimer population rotation.
 The  transition operators are applied by concatenated implementation of intra- and inter-dimer population swapping  i.e.\\
$T_{xyz}=\text{e}^{\text{i}H_{xyz1}\pi/2}R(\pi/2)$, $T_{xy}=\text{e}^{\text{i}H_{xy1}\pi/2}R(\pi/2)$, $T_{xz}=\text{e}^{\text{i}H_{xz1}\pi/2}R(\pi/2)$, $T_{x}=\text{e}^{\text{i}H_{x1}\pi/2}R(\pi/2)$, $T_{y}=\text{e}^{\text{i}H_{xy0}\pi/2}R(\pi/2)$,  $T_{z}=\text{e}^{\text{i}H_{xz0}\pi/2}R(\pi/2)$. Extended forms of transition operators are presented in Eq.~\ref{Eq_T3DCoined}.

\begin{figure}
\centering
\scalebox{0.4}{\includegraphics*{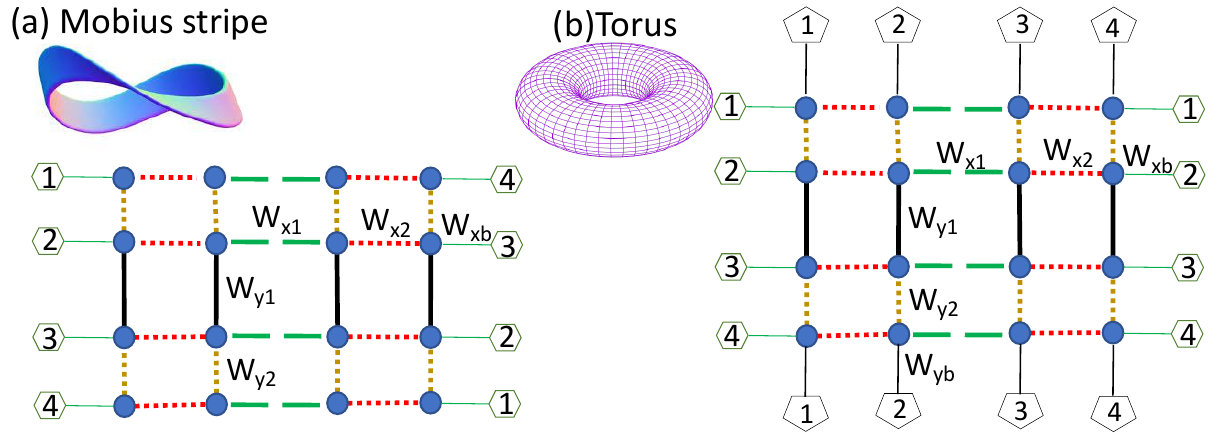}}
\caption{Rydberg DTQW over topological surfaces  (a) M\"obius stripe and (b) torus. To realize the boundary conditions the pair  sites with the same pentagon (hexagon) number would get excited to $nS$ state with local lasers over the boundary operation step $W_{yb}$ ($W_{xb}$). Concatenated operation of $W_{y0}W_{y1}W_{yb}W_{x0}W_{x1}W_{xb}$, performs the QW on the desired topological surface \cite{Moq18}.}\label{STEPS}
\end{figure}

Quantum walk over {\it topological periodic boundary conditions} would enrich the physics \cite{Mich17,Moq18} and diverse the possible applications \cite{Kit10,Xia10,Gru14,Lac16}.
The long-range Rydberg interaction could be used for realizing DTQW with topological periodic boundary conditions on an atomic lattice. Fig.~\ref{STEPS} shows two examples of DTQW over (a) M\"obius stripe and (b) torus  topological surfaces. While the torus boundary condition could be implemented by global laser over limited lattice sites, forming other topological surfaces, e.g.  M\"obius and Kline bottle, requires local laser excitations. During the boundary operation step $W_{yb}$ ($W_{xb}$) with the local lasers, the pair sites with the same pentagon (hexagon) numbers will be excited to $\ket{s}$ under local lasers with detuning adjusted to the exchange interaction. The sequence of the QW operators  will be $U=W_{y0}W_{y1}W_{yb}W_{x0}W_{x1}W_{xb}$. 
 
\section{Decoherence in Rydberg DTQW}
\label{Sec_Decoherence}

\subsection{Non-unitary dynamics}
\begin{figure}
\centering
 \subfloat{%
  \includegraphics[width=.162\textwidth]{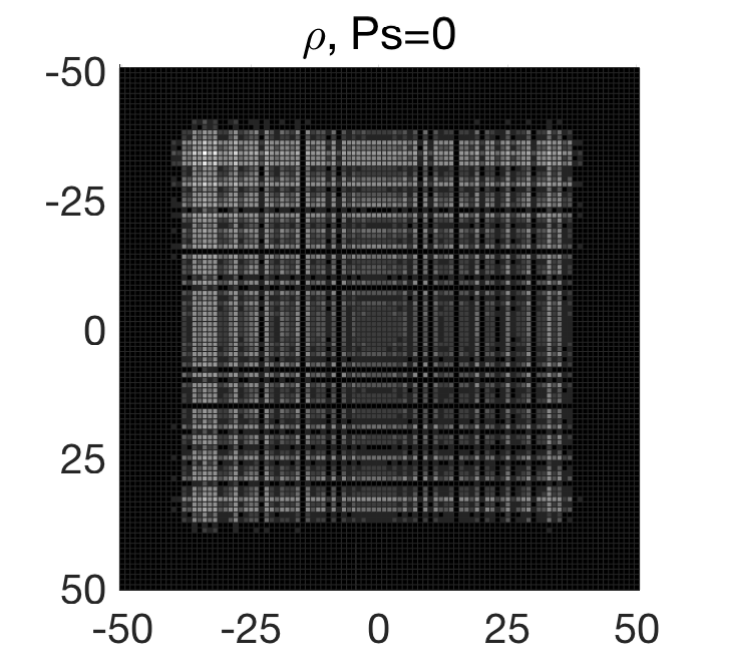}}\hfill 
 \subfloat{%
   \includegraphics[width=.155\textwidth]{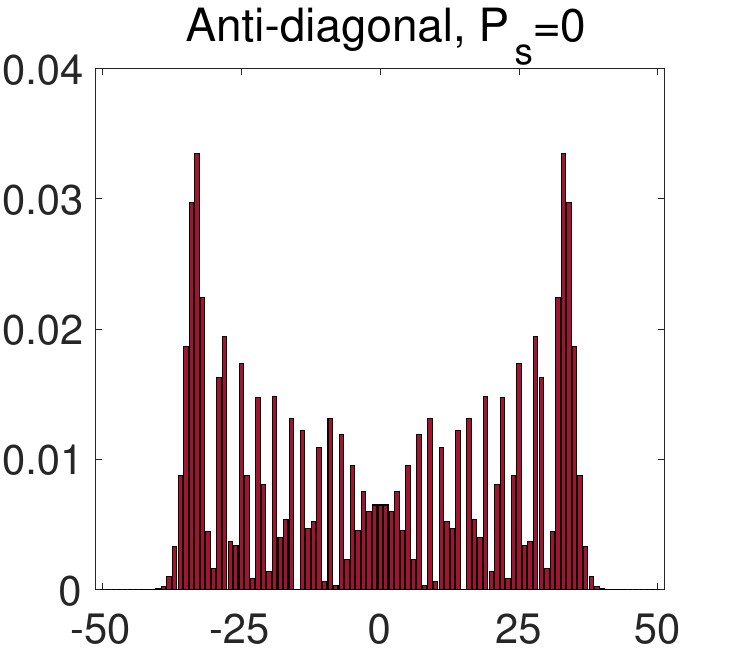}}\hfill  
   \subfloat{%
   \includegraphics[width=.155\textwidth]{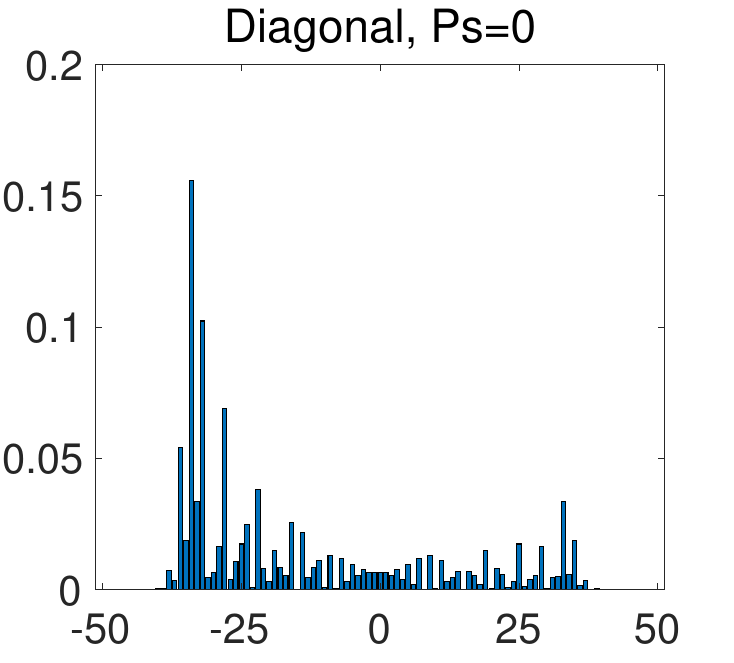}}\hfill  
    \subfloat{%
  \includegraphics[width=.162\textwidth]{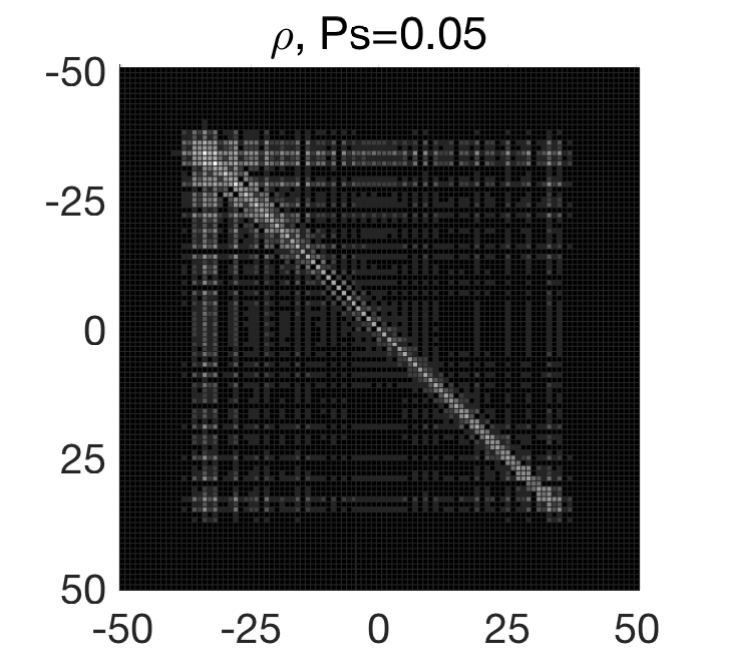}}\hfill 
 \subfloat{%
   \includegraphics[width=.155\textwidth]{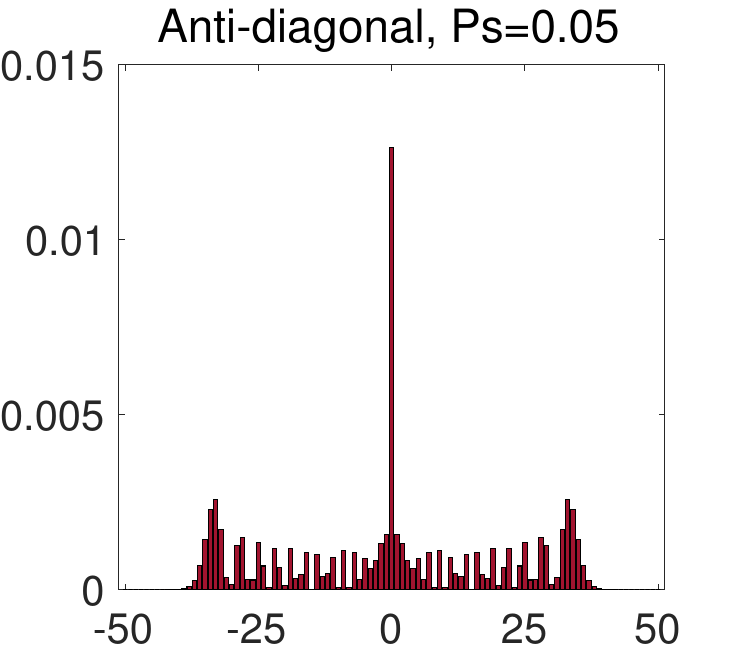}}\hfill  
   \subfloat{%
   \includegraphics[width=.155\textwidth]{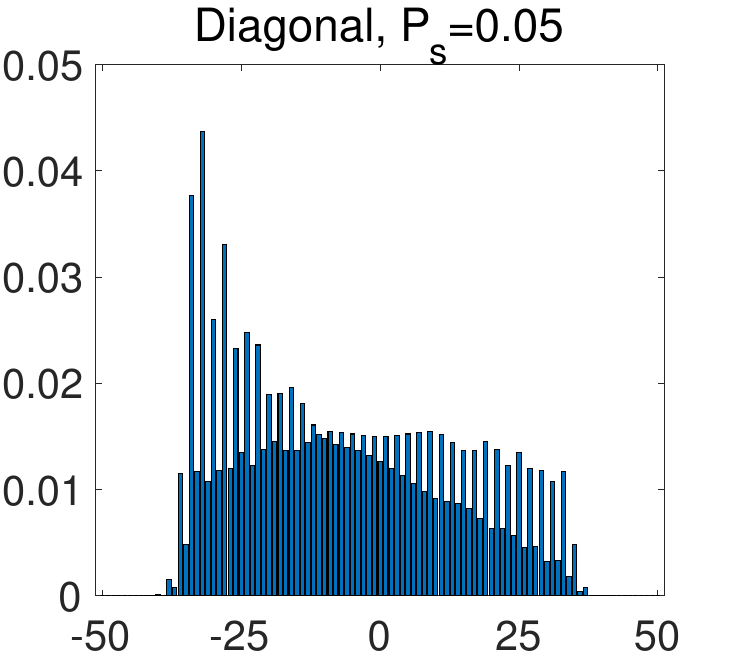}}\hfill  
    \subfloat{%
  \includegraphics[width=.162\textwidth]{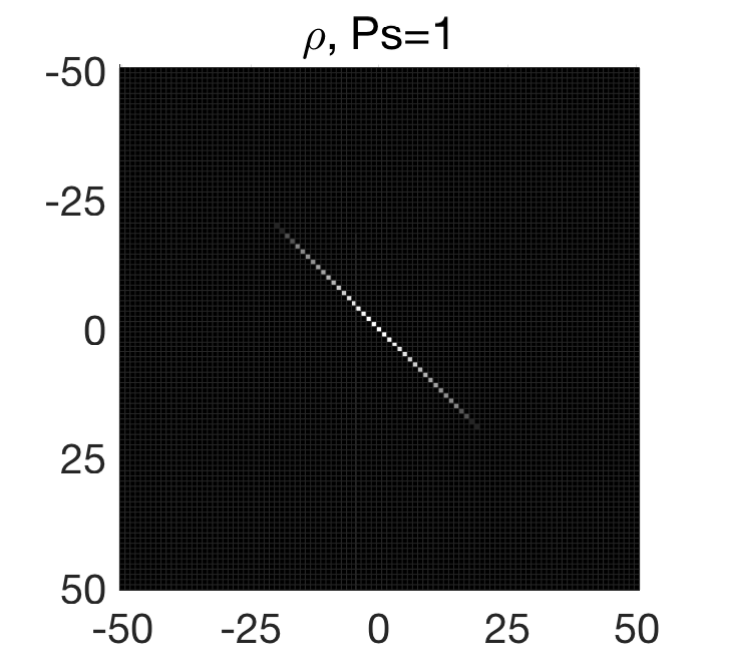}}\hfill 
 \subfloat{%
   \includegraphics[width=.155\textwidth]{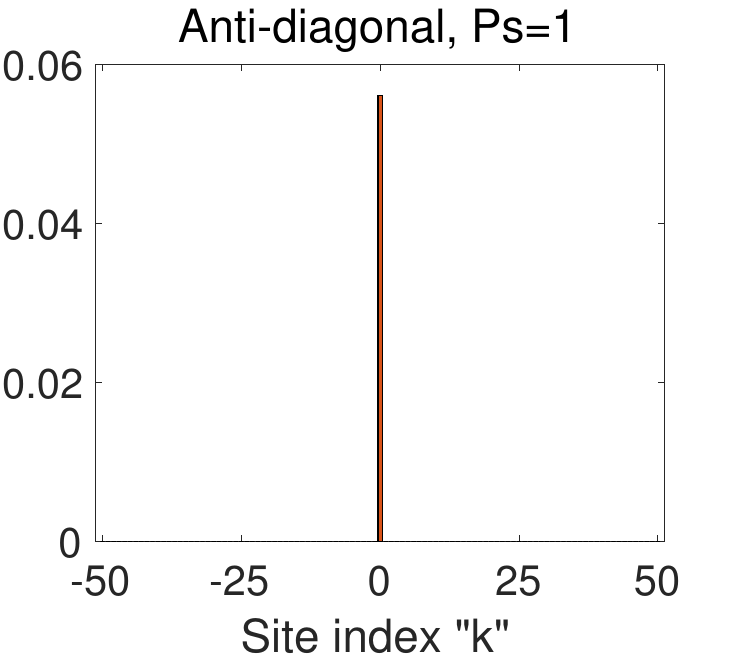}}\hfill  
   \subfloat{%
   \includegraphics[width=.155\textwidth]{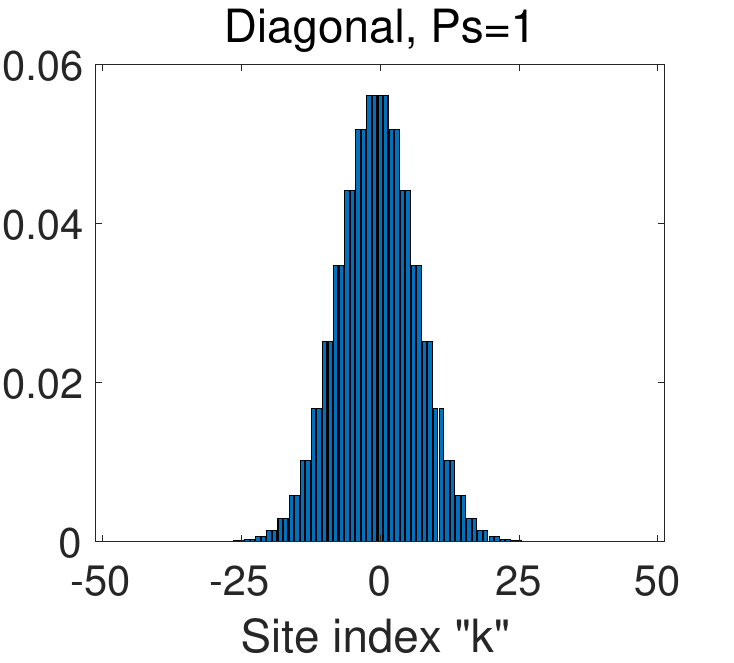}}\hfill  
\caption{Effects of de-phasing on (left column) density matrix terms $\rho(x,x')$, (middle column) anti-diagonal coherence $\rho(x,-x)$ and (right column)  probability distribution $\rho(x,x)$ after 50 Hadamard steps $\theta=\pi/4$ of coinless Rydberg DTQW over a 1D lattice with N=101 sites. The three rows from top to bottom are corresponding to $P_s=[0, 0.05, 1]$ de-phasing probability over single step.}\label{Fig_DensityCoh}
\end{figure}

 The non-unitary dynamics of the quantum walk can be described by the projection of quantum state onto the pointer states \cite{Schl07,Alb14}. 
In this model, the pointer state projector $\Pi_{x}= \ket{x}\bra{ x}$, projects the walker  into the $x^\text{th}$ site.
In the presented Rydberg QW model, the evolution of quantum walker over a single step is mainly coherent. Hence, the decoherence could be applied in a discrete-time manner after each step. The effective stroboscopic non-unitary operator would be
\begin{equation}
\rho_{i+1}=(1-P_s)W\rho_iW^{\dagger}+P_s\sum\limits_{x} \Pi_{x} W \rho_i W^{\dagger} \Pi_{x}^{\dagger}
\end{equation} 
where $\rho_i=\sum_{x,x'} \ket{x}\bra{ x'}$ is the density operator after the i$^\text{th}$ step.
The spatial de-phasing terms discussed in the next section would reduce the off-diagonal coherence terms with a probability of $P_s$ after each step and hence suppress the superposition effect. In the totally diffusive case $P_s=1$, the absence of the interference of quantum paths would only leave a classical Gaussian probability distribution in the diagonal terms, see Fig.~\ref{Fig_DensityCoh}.

\begin{figure}
\centering
 \subfloat{%
  \includegraphics[width=.235\textwidth]{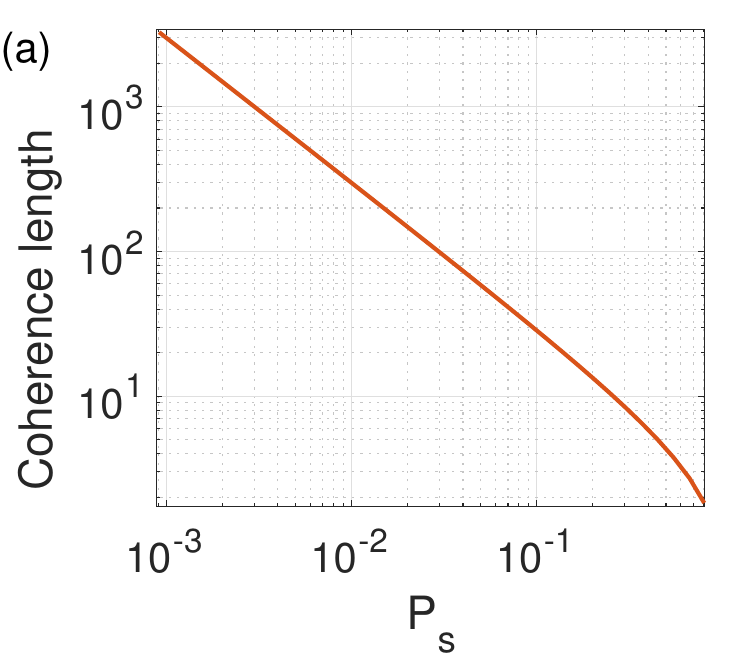}}\hfill 
 \subfloat{%
   \includegraphics[width=.235\textwidth]{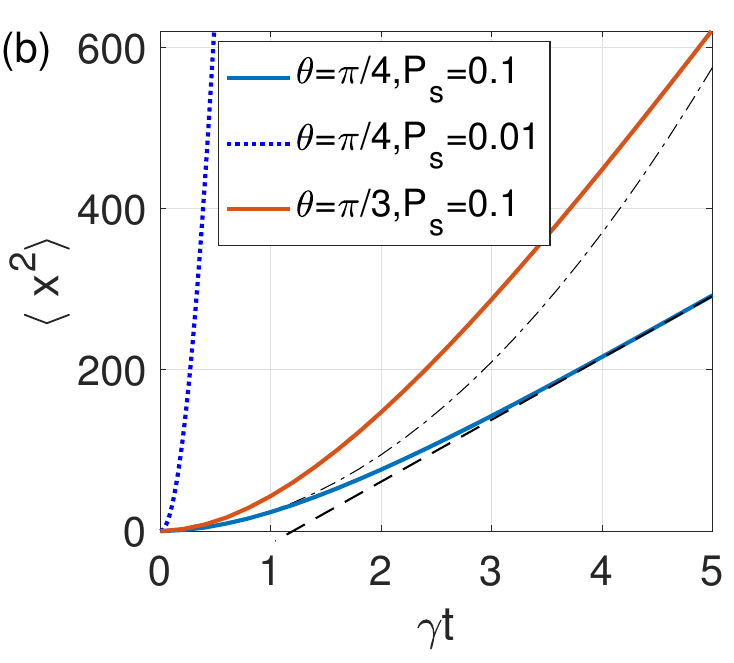}}\hfill  
\caption{Effects of dephasing on (a) coherent-length and (b) Ballistic distribution. (a) The number of lattice-sites over which the coherence preserves is numerically calculated, see the text for more details.  Figure (b),  numerically calculates $<x^2>$ after 50 steps in N=101 lattice dimension as a function of $\gamma t=i P_s$ where $i$ is the step number, $\gamma$ is the total dephasing rate and $t$ is the operation time.
For time scales below $\gamma^{-1}$ the square displacement is ballistic $\propto t^2$ while above that it becomes diffusive $\propto t$.  Dotted-dashed line is a quadratic fit to the ballistic part $22.5(\gamma t)^2$. The hopping angle in (a) is $\theta=\frac{\pi}{4}$.  }\label{Fig_CohLenBal}
\end{figure}

To define the coherence length $l_{co}$ one can look at the suppression of anti-diagonal terms $|\rho_{(x,-x)}|=|\rho_{_0(x,-x)}|\exp(-|x|/l_{co})$ with respect to the unperturbed one $\rho_{_0(x,-x)}$. Coherence length is plotted as a function of $P_s$ in Fig.~\ref{Fig_CohLenBal}a. The crossover between coherent and incoherent cases can also be seen in the rate of mean square displacement $ \langle x^2\rangle =\sum_{x} \rho(x,x)x^2$.
Fig.~\ref{Fig_CohLenBal}b, shows that within the coherence period $t<\Gamma^{-1}$ the walker propagates ballistically  $\propto t^2$ while in the diffusive regime $t>\Gamma^{-1}$  it propagates linearly with time.    
Overall, Fig.~\ref{Fig_CohLenBal}, shows that  coherent staggered quantum walk over a lattice with $N\approx P_s^{-1}$ sites propagates ballistically over $i=P_s^{-1}$ steps. In the next section, $P_s$ is evaluated in the proposed scheme.

\subsection{Sources of errors}
\label{Sec_Error}
\subsubsection*{Laser error sources}
{\it Laser noise} over the Rydberg excitation process, causes de-phasing. The laser noise $\gamma_{la}$ that is encountered here is what remains after laser locking in the two-photon excitation.  
Fig.~\ref{Fig_DephasSour}a, simulates the de-phasing probability $P_s$ after one step as a function of the relative laser noise $\frac{\gamma_{la}}{\Omega}$. In this figure,  effects of the laser-noise over the Quantum walk operation is simulated using master equation with  $\sqrt{\gamma_{\text{la}}} \ket{s}\bra{ s}$ Lindblad term.
In the recent experiments  \cite{Ber17,Lev18}, cavity phase noise filtering  suppresses the laser noise below the effective bandwidth
of the lock, resulting in narrow line-widths of  0.5 kHz for the two-photon Rabi frequency of $\Omega/2\pi=$2MHz. Thus, dephasing probability over one  Hadamard step ($\theta=\frac{\pi}{4}$) is  $P_{s}\approx10^{-4}$. 

{\it Spontaneous scattering} from the optical lattice lasers as well as  Rydberg exciting lasers,  destroy the coherence by projecting the quantum walker's wave-function into a single lattice site. 
The new advances in clock optical lattices have  suppressed the trap lasers' scattering rate, reaching  coherence times of 140s \cite{Ros19} for lattice constants above 2$\mu$m, making the corresponding dephasing per step negligible $P_s\approx10^{-8}$.
Spontaneous emission also occurs from the intermediate state $ \ket{p}$, over the two-photon Rydberg excitation $\ket{g}\bra{s}$. The two lasers $\Omega_1$, $\Omega_2$ are detuned from the intermediate level by $\Delta_p$.
The dominant decay channel from $ \ket{p}$ is back into the ground state $ \ket{g}$. This would result in an effective Lindblad de-phasing term $\sqrt{\gamma_p}\ket{g} \bra{ g}$, where $\gamma_{p}/2\pi=1.4$MHz  is decay rate of the intermediate level $ \ket{p}=6P_{3/2}$ in Rb.
Over one quantum step operation time of $\frac{2\pi}{\Omega}$ with effective  Rabi frequency of $\Omega=\Omega_1\Omega_2/2\Delta_{p}$, the de-phasing probability after one step would be  $P_s=\frac{\pi \gamma_{p}}{2\Delta_{p}}(\frac{\Omega_1}{\Omega_2}+\frac{\Omega_2}{\Omega_1})$ \cite{Saf10}. 
Using the   experiment parameters in exciting $100S$ via $ \ket{p}=6P_{3/2}$ intermediate level \cite{Lev18} with $(\Omega_1,\Omega_2)=2\pi\times (60,40)$MHz for (420nm,1013nm) lasers and the detuning of $\Delta_p/2\pi=600$MHz the effective Rabi frequency would be $\Omega/2\pi=2$MHz and de-phasing probability over single quantum step would be $P_{s}=2.5\times10^{-4}$.

\subsubsection*{Lattice geometry and confinement}

\begin{figure}
\centering
 \subfloat[]{%
   \includegraphics[width=.23\textwidth]{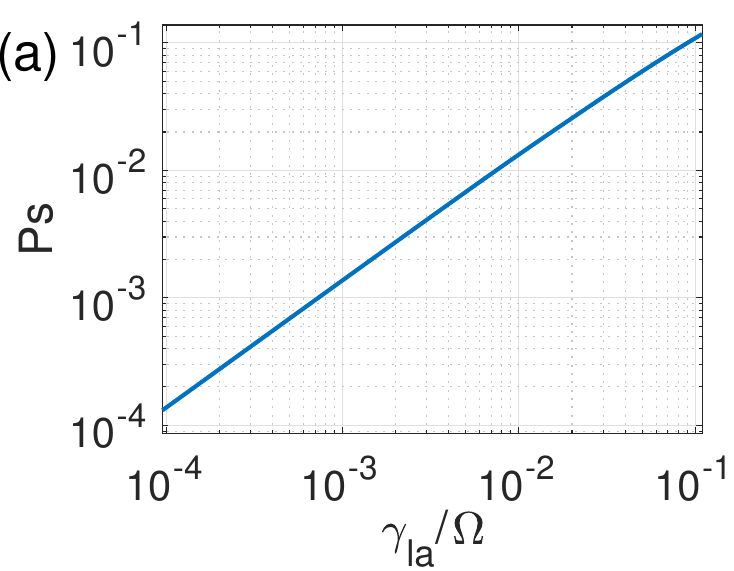}}\hfill  
    \subfloat[]{%
   \includegraphics[width=.23\textwidth]{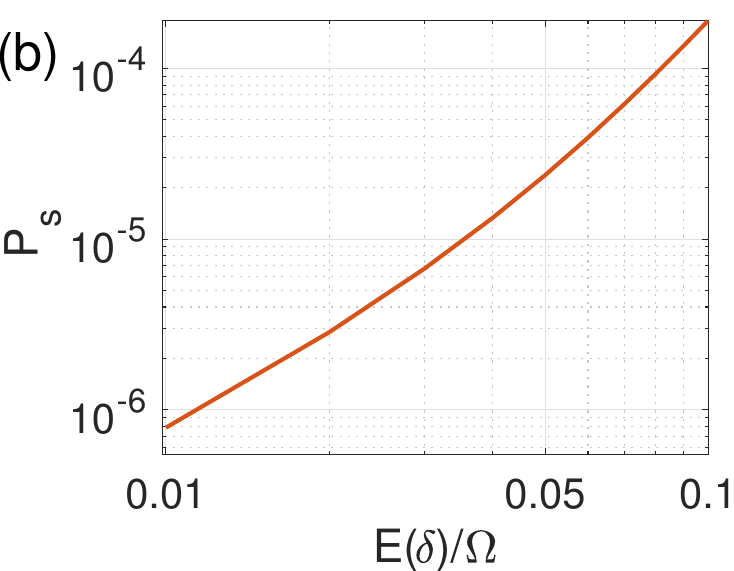}}\hfill  
\caption{  Single-step de-phasing probability $P_s$  over a Hadamard step ($\theta=\pi/4$) as a function of (a)  relative laser noise $\gamma_{\text{la}}/\Omega$ and (b) transition detuning errors $\frac{E(\delta)}{\Omega}$.}\label{Fig_DephasSour}
\end{figure}

The known detuning would cause a phase that gets absorbed in the definition of $ \ket{g_k}$ state. However, the random fluctuations of detuning $E(\delta)$ caused by spatial uncertainty and Doppler broadening leads to spatial de-phasing (see Fig.~\ref{Fig_DephasSour}b) and could affect the population rotation designed for the quantum jump protocol. 

{\it Confinement:} 
 The interaction variation over the spatial profile of trapped atoms, results to an uncertainty of the laser detuning $E_{\delta}$. 
Considering 6$\mu$m and 9$\mu$m lattice constant, and the motional ground state expansion of FWHM=25nm for Rb atoms with trap frequency of  $\omega_{tr}/2\pi= 165$kHz \cite{Wan19}, the root mean square deviation of detuning would be  $E_{\delta}/\Omega=0.03$ and 0.007 for principal number $n$=100 and $\Omega/2\pi=2$MHz. This would cause $P_s= 7\times10^{-6}$ and $3\times 10^{-7}$  de-phasing over a Hadamard  step, see Fig.  \ref{Fig_DephasSour}b.  Further confinement of atoms could be obtained using the dark-state optical lattices \cite{Yav09}.

{\it Doppler broadening:} 
Detuning error could also be caused by Doppler broadening. Using counter-propagating 1013nm and 420nm lasers for two-photon Rydberg excitation, the Doppler detuning would be $\delta_D=(k_1-k_2)v$.
The atomic motion could be suppressed by sideband cooling within the optical tweezer \cite{Kau12,Tho13} and optical lattice \cite{Bel13} to the motional ground state.
At the ground motional state, the maximum velocity of the thermal motion  would be $v=\sqrt{\hbar\omega_{tr}/2m}=18$mm/s. 
The random Doppler shift generates a maximum relative uncertainty of $\frac{\delta_D}{\Omega}=0.03$ for  $\Omega/2\pi=2$MHz. 
The corresponding de-phasing probability over a Hadamrd step is $P_s=3\times10^{-6}$, see Fig.~\ref{Fig_DephasSour}b. 

{\it Spontaneous emission--}
Loss out of the system from the Rydberg levels $S$ and $P$, reduces  both diagonal and off-diagonal terms of density matrix alike \cite{Wus11}. Hence, after the projective measurement, the loss does not effect the coherence of the QW operation and would only cause an overall decay, as quantified in Sec.~\ref{Sec_Fid}. A small portion of the $100P$ state decays to the ground state with 110s$^{-1}$ rate. Over a quantum step performed by $\Omega/2\pi=2$MHz the  dephasing probability would be $P_s=5\times10^{-5}$.

\section{Operation Fidelity in 3D lattice}
\label{Sec_Fidelity3D}
This section defines the QW fidelity in terms of the accuracy of population transfer to the desired site with correct hopping angle and does not encounter the decoherence effects discussed in the previous section.
The implementation of 3D QW with coupling tessellations presented in Fig.~\ref{Fig_MultiD}b and Eq.~\ref{Eq_H3d} benefits from the interaction's angular dependency.
After overviewing the anisotropic exchange interaction, the optimum lattice constant and the operation fidelity of different coupling tessellations are quantified. Finally, the scaling of achievable step numbers are discussed as a function of applied principal numbers.

\subsection{ Angular-dependent interaction}
\label{Sec_Int}
\begin{figure}
\centering
\scalebox{0.48}{\includegraphics*[viewport=25 20 700 230]{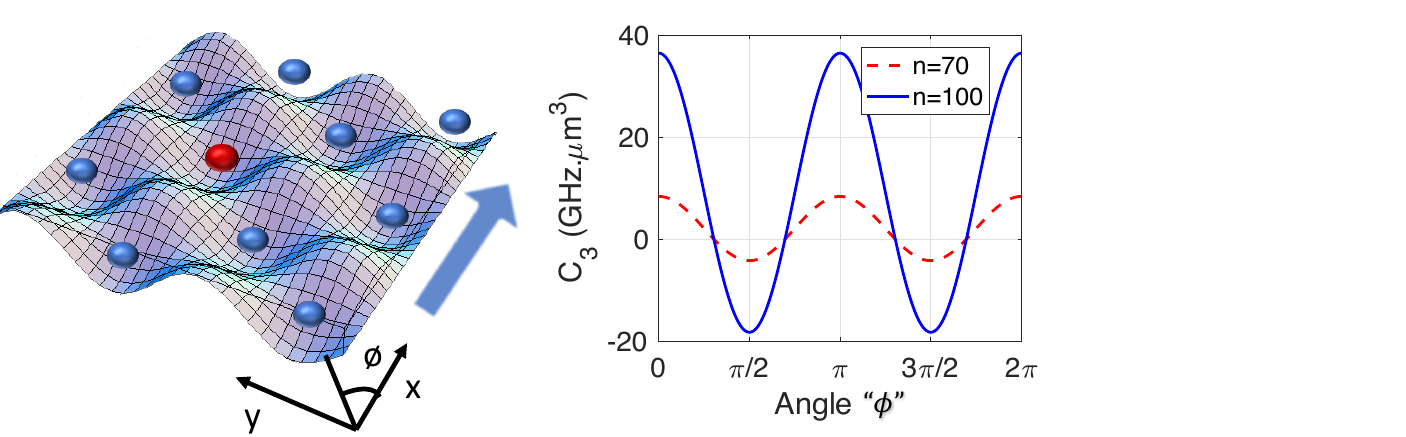}}
\caption{The anisotropic $nS-nP$ exchange interaction $V$  provides a wider range of laser detuning in both positive and negative sides, available for addressing the desired site. Here $\phi$ is the angle between the inter-atomic orientation and the quantization axis.  Quantization axis is defined by the propagation direction of the polarized exciting laser.
}\label{Fig_Anisotropic}
\end{figure}

The angular-dependent exchange interaction of  $\ket{nS_{1/2}1/2, \, nP_{3/2}3/2} {\stackrel{V}{\rightleftharpoons}} \ket{nP_{3/2}3/2, \, nS_{1/2}1/2}$ is given by
\begin{equation}
    V=\frac{ \bm{\mu}_1.\bm{\mu}_2-3(\bm{\mu}_1.\hat{R})(\bm{\mu}_2.\hat{R})}{4 \pi \epsilon_0 R^3}
\end{equation}
where $R$ is the interatomic separation and $\vec{\mu_k}$ is the electric dipole matrix element, connecting initial and final Rydberg state of $k^{th}$ atom. 
The angular dependent interaction between sites $i$ and $j$ could be expanded to
\begin{eqnarray}\nonumber
\label{Eq_Vij}
&&V_{ij}(\phi)=\frac{1}{4\pi \epsilon_0 R_{ij}^3}\\
&&[f_1(\phi)(\mu_{1+}\mu_{2-}+\mu_{1-}\mu_{2+}+2\mu_{1z}\mu_{2z})\\ \nonumber
&&+f_2(\phi)(\mu_{1+}\mu_{2z}-\mu_{1-}\mu_{2z}+\mu_{1z}\mu_{2+}-\mu_{1z}\mu_{2-})\\ \nonumber
&&-f_3(\phi)(\mu_{1+}\mu_{2+}-\mu_{1-}\mu_{2-})]=\frac{C_3(\phi)}{R_{ij}^3},\nonumber
\end{eqnarray}
where $\phi$ shows inter-atomic orientation with respect to the quantization axis, defined by the propagating direction of exciting polarized lasers. Dipole operators in the spherical basis are denoted by $\mu_{k,\pm}=\mp(\mu_{k,x}\pm i\mu_{k,y})/\sqrt{2}$. The terms associated with pre-factors $f_1(\phi)=(1-3\cos^2\phi)/2$, $f_2(\phi)=\frac{3}{\sqrt{2}}\sin \phi \cos \phi$ and $f_3(\phi)=3/2 \sin^2 \phi$ couple Rydberg pairs. 
Fig.~\ref{Fig_Anisotropic}b represents the angular dependent exchange interaction for two principal numbers.

\subsection{Optimum lattice constant}

\begin{figure}
\centering
\scalebox{0.45}{\includegraphics{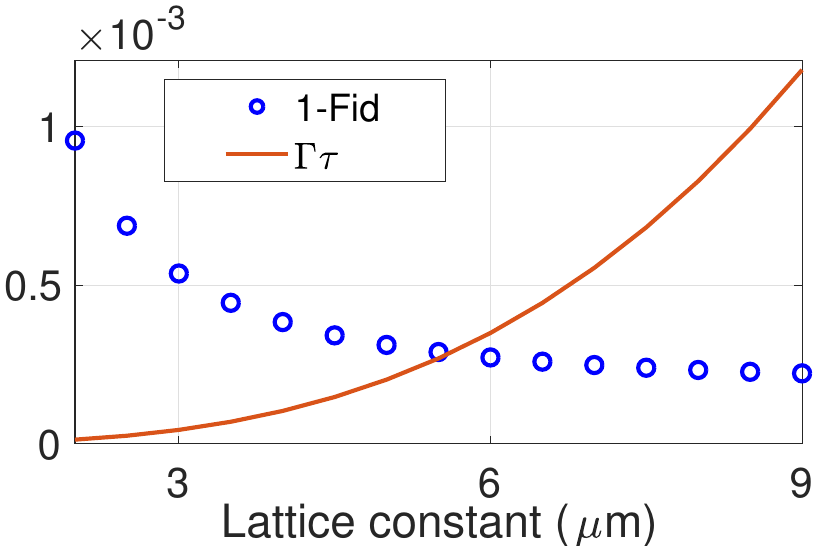}}
\caption{ Optimum lattice constant. Smaller lattice constants $a$ results in stronger interaction.  Hence, faster operation ($\Omega/V(a)=0.1$, n=100) reduces the spontaneous emission over each step $\Gamma \tau$, where $\Gamma$ is the decay rate of 100P state at the cryogenic environment T=77K and $\tau$ is the operation time. Considering tight confinement of atoms in ground motional state with FWHM=25nm, the interatomic uncertainty increases the average infidelity of quantum walk (plotted here for $\theta=\pi/2$) at smaller lattice constants.
}\label{Fig_Confignment}
\end{figure}

Smaller lattice constants results in stronger interaction and hence faster operation. In the other hand, it enhances the Rydberg pair level-mixing as well as the errors associated with the uncertainties in the inter-atomic distance.
 Here, tightly confined traps are considered, where the ground motional state of the Wannier function at each site could be represented by a Gaussian profile with the FWHM of 25nm. The corresponding uncertainty in the relative inter-atomic distance would raise an average error in population transfer as shown by circles in Fig.~\ref{Fig_Confignment}. The lattice constant range of $6\mu$m$<a<9\mu$m, results in high precision operation while keeping the spontaneous emission of the Rydberg level $\Gamma\tau$ low, see Fig.~\ref{Fig_Confignment}.
Considering the sensitivity of the proposed scheme to interatomic distance, simultaneous confinement of both Rydberg and ground state atoms is important for high fidelity operation. This has been realized in both alkaline \cite{Zha11,Bar20,Gra19} and alkaline earth atoms \cite{Coo18,Wil19}. 
 
 The Rydberg pair level-mixing at $n=100$, could safely be neglected at  6$\mu$m and $9\mu$m lattice constants isolating the  desired exchange channel $SP {\stackrel{V}{\rightleftharpoons}} PS$.
 The next strongest off resonant coupled Rydberg pair $\ket{100S_{1/2}1/2, \, 100P_{3/2}3/2} {\stackrel{V_2}{\rightleftharpoons}} \ket{99P_{3/2}3/2, \, 99D_{5/2}5/2}$  would get negligible population of $(V_2/\delta_E)^2\approx10^{-4}$ and $10^{-5}$ over the process, where $\delta_E$ is the energy difference of the two pairs.   
 The relative perturbation of the desired interaction $V$  would be of the order  $10^{-5}$ and $10^{-6}$.
 Consequently, off-resonant pair channels do not affect the laser population rotation in the scheme and are safely neglected.

The other potential source of error is due to the off-resonant excitation of nS Rydberg atoms over the entire lattice.  At the considered regime of operation $\Omega/\Delta=0.1$, the population leakage of each site in the lattice would be  $P_{\text{leak}}=(\Omega/\Delta)^2=0.01$.
Considering the scale of van-der-Waals interaction among $nS-nS$ atoms as $V_{SS}\propto r^{-6}$, a cut off after the nearest neighbouring sites could be considered. As a result, the cross-talk among the leaked $\ket{nS}$ population, does not scale with the total lattice site number.  
In a 3D lattice of dimers with $a_{\{x_0,x_1,y,z\}}=(6,9,6,6)\mu$m  lattice constants the level-shift  caused by the global population leakage of $100S$  would be four orders of magnitude smaller than $\Omega$ and hence does not effect the laser's fine tuning.

\subsection{ Coupling tessellations' fine tuning in 3D}
\label{Sec_Fid}

\begin{figure}
\centering
 \subfloat{%
  \includegraphics[width=.237\textwidth]{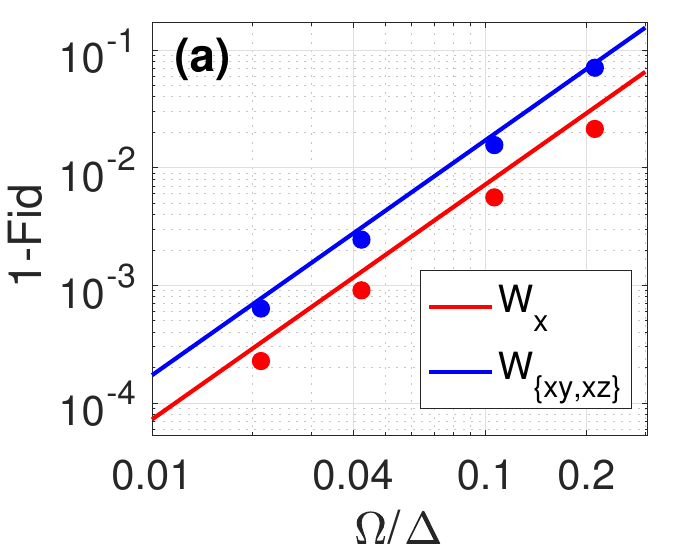}}\hfill 
 \subfloat{%
   \includegraphics[width=.237\textwidth]{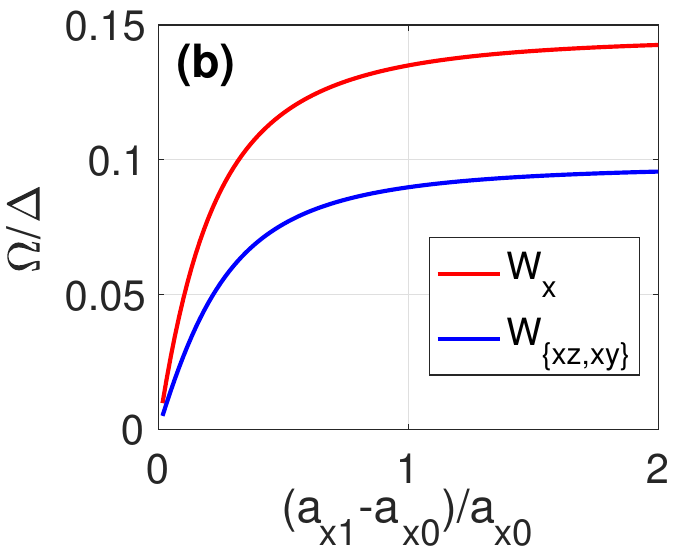}}\hfill  
\caption{Fidelity of a spin transport ($\theta=\pi/2$) in a  3D lattice of dimers of Fig.~\ref{Fig_MultiD}b. (a) Infidelity of desired coupling tessellations  are plotted as a function of $\Omega/\Delta$ calculated by the analytical model (lines) and numerical simulation (circles). The two methods considers square and Gaussian laser pulses respectively and encounters 8 neighbouring lattice sites.  (b) The required relative laser coupling $\Omega/\Delta$ is plotted as a function of the contrast of inter- and intra-dimer lattice constants $(a_{x0}-a_{x1})/a_{x0}$ to realize connectivities with 99\% fidelity.  The lattice constants are $a_{x0}=a_y=a_z=6\mu$m in (a,b) and  $a_{x1}=9\mu$m in (a). }\label{Fig_Fidelity}
\end{figure}

This section quantifies the QW infidelity, caused by population scattering to the undesired sites in a 3D lattice structure. 
Tuning the laser to be in resonance with the interacting sites $i$ and $j$ separated by ${\bf R}_{ij}$ i.e.  $\Delta=-V_{ij}$, the operation infidelity would be the sum of population leakage to unwanted sites $k$ i.e. 
\begin{equation}
\label{Eq_Fid}
1-F_{ij}=\sum\limits_{k\neq i,j}\frac{\Omega^2}{4\delta_k^2}.
\end{equation} 
where the effective laser detuning in site $k$ is given by $\delta_k=\Delta+V_{ik}$. 
Since the laser is oriented along $R_{ij}$, $\delta_k$ would be the bare atom detuning $\Delta=-C_3(0)/R_{ij}^3$  modified by the interaction with the quantum walker $V_{ik}=C_3(\phi_k)/R_{ik}^3$ with $\phi_k$ being the angle between  ${\bf R}_{ij}$ and ${\bf R}_{ik}$.
The infidelity of desired coupling tessellations of Fig.~\ref{Fig_MultiD}b are plotted in Fig.~\ref{Fig_Fidelity}a as a function of $\Omega/\Delta$. The lines are presenting the infidelity based on the analytical approach of Eq.~\ref{Eq_Fid} for a square laser pulse, while the circle signs calculates the infidelity based on the numerical simulation of the Schr\"odinger equation Eq.~\ref{Eq_HRy} with a Gaussian pulse $\Omega(t)=\Omega e^{-\frac{(t-\tau/2)^2}{2\sigma^2}}e^{-\frac{(\tau/2)^2}{2\sigma^2}}$
where  $\sigma=\tau/5$ and a pulse duration $\tau$ given by $\int_0^{\tau}\Omega(t)\text{d}t=2\pi$. The calculations are encountering  8 neighboring lattice sites at different inetr-atomic orientations.   
The contrast of inter- and intra-dimer lattice constants $(a_{x0}-a_{x1})/a_{x0}$ would define the speed of operation. In general,  better contrast leads to faster operation for a given fidelity, see Fig.~\ref{Fig_Fidelity}b. 

Realizable {\it QW step number} scales  with the Rydberg principal number $n$.
 While the loss of population from the Rydberg state has minor effect on the coherence (see Sec.~\ref{Sec_Error}), it determines the possible step numbers.
 In this sense, faster operation at constant $\Omega/\Delta$ would require stronger interaction $\Delta=-C_3/a^3$. While $C_3\propto n^4$, the minimum lattice constant is limited by the LeRoy-radius and hence scales as  $a\propto n^2$. 
The loss  over a single QW step thus  is scaled by  $2\pi\Gamma/\Omega \propto n^{-1}$ where  $\Gamma\propto n^{-3}$ \cite{Bet09} is the   spontaneous emission rate of Rydberg state.
The step numbers would be significantly enhanced by coherent fast excitation of circular states \cite{Sig17,Car20}  featuring exchange interaction and several minutes lifetime \cite{Ngu18}.

Finally, the {\it initialization and detection} must be taking into account while quantifying the device operation. After preselection, the probability of correct state initialization  of more than 98\% has been achieved \cite{Ber17}.
Fluorescence detection of ground-state atoms has been realized with 98\% fidelity in Rb \cite{Kwo17} and 0.99991 fidelity in Sr \cite{Cov19}.

\section{Discussion and Outlook}
This article, shows that smart Rydberg excitation in a holographically designed atomic lattice act as a versatile platform for multi-dimensional DTQW.  
The proposed model could be used for making an enriched form of  Floquet topological insulators, as formulated and discussed in the appendix A2. 
The robustness of the Floquet topological insulator against fluctuations of the hopping angles, allows fast operations with lower fidelity requirement, see Fig. \ref{Fig_Confignment}, \ref{Fig_Fidelity}.  This flexibility would significantly enhance the number of coherent QW steps over larger lattices making the Rydberg proposal an ideal platform for large-scale  simulation of multi-dimensional topological insulators.

In the other direction, the presented model could be used for the simulation of electron movement on the Fermi surfaces that are 2D manifolds fitted in the Brillouin zone of a crystal (electron movement on the Fermi surfaces that cut each other like Kline bottle surface) \cite{Wie16,Moq18}. The other extension avenue would be obtained by adding synthetic magnetic fields to the current DTQW model. This would be obtained by applying a gradient of an external electric field resulting in magnetic QW with applications in making Chern insulators \cite{saj18}.
The new features of all to all connectivity, provide a platform to study the performance of QW based search algorithms \cite{Nei03} on topologically ordered spatial databases, along the computer science studies \cite{Seg00,Cle02}.

\section*{Supplemental material}

\appendix

 \section*{A1: Expanded form of the Rydberg DTQW operators in multi-dimensions}  
 
The operators and Hamiltonians of Rydberg multi-dimensional DTQW are presented in expanded forms in this appendix. The coupling Hamiltonians in a 2D lattice of tetramers plotted in Fig.~\ref{Fig_SupFig2}a are 
\begin{eqnarray}
\label{Eq_2DTetramers}
&&H_{x0}=\sum\limits_{m_x=1}^{N_x/2} (\ket{m_{x},e_x} \bra{ m_x,o_x}\otimes \mathbbm{1}_y+\text{h.c.}) \\ \nonumber
&&H_{x1}=\sum\limits_{m_x=1}^{N_x/2} (\ket{m_x,e_x} \bra{ m_x+1,o_x}\otimes \mathbbm{1}_y+\text{h.c.})\\ \nonumber
&&H_{y0}=\sum\limits_{m_y=1}^{N_y/2} ( \mathbbm{1}_x \otimes \ket{m_y,e_y} \bra{ m_y,o_y}+\text{h.c.}) \\ \nonumber
&&H_{y1}=\sum\limits_{m_y=1}^{N_y/2} ( \mathbbm{1}_x \otimes \ket{m_y,e_y} \bra{ (m_y+1),o_y }+\text{h.c.})\\ \nonumber
\end{eqnarray} 
where the sub-lattice elements are consists of distinguished odd and even sites in each dimension.
 The coupling Hamiltonians in a 3D lattice of dimers plotted in Fig.~2b and \ref{Fig_SupFig2}c are
 
 \begin{eqnarray}
\label{Eq_H3d}
&&H_{x0}=\sum\limits_{{\bf m}} (\ket{m_{x},m_y,m_z,e} \bra{ m_x,m_y,m_z,o}+\text{h.c.}) \\ \nonumber
&&H_{x1}=\sum\limits_{{\bf m}} (\ket{m_x,m_y,m_z,e} \bra{ m_x+1,m_y,m_z,o}+\text{h.c.})\\ \nonumber
&&H_{xy0}=\sum\limits_{{\bf m}} ( \ket{m_x,m_y,m_z,e} \bra{ m_x,m_y+1,m_z,o }+\text{h.c.})\\ \nonumber
&&H_{xy1}=\sum\limits_{{\bf m}} (  \ket{m_x,m_y+1,m_z,e} \bra{ m_x+1,m_y,m_z,o }+\text{h.c.})\\ \nonumber
&&H_{xz0}=\sum\limits_{{\bf m}} ( \ket{m_x,m_y,m_z+1,e} \bra{ m_x,m_y,m_z,o }+\text{h.c.})\\ \nonumber
&&H_{xz1}=\sum\limits_{{\bf m}} (  \ket{m_x,m_y,m_z,e} \bra{ m_x+1,m_y,m_z+1,o }+\text{h.c.})\\ \nonumber
&&H_{xyz1}=\sum\limits_{{\bf m}}   \ket{m_x,m_y,m_z,e} \bra{ m_x+1,m_y+1,m_z+1,o }+\text{hc}\\ \nonumber
\end{eqnarray} 

In the proposed coined DTQW, the coin rotation operator is applied by the intra-dimer population rotation.   
The  transition operators are applied by the concatenated implementation of intra- and inter-dimer population swapping  i.e.\\
\begin{widetext}
\begin{center}
\begin{tabular}{ c c  c }
$T_{x}=\text{e}^{\text{i}H_{x1}\pi/2}R(\pi/2)$ \qquad & \qquad $T_{y}=\text{e}^{\text{i}H_{xy0}\pi/2}R(\pi/2)$ \qquad &    \qquad $T_{z}=\text{e}^{\text{i}H_{xz0}\pi/2}R(\pi/2)$\\
 $T_{xyz}=\text{e}^{\text{i}H_{xyz1}\pi/2}R(\pi/2)$ \qquad & \qquad $T_{xy}=\text{e}^{\text{i}H_{xy1}\pi/2}R(\pi/2)$	 \qquad &\qquad     $T_{xz}=\text{e}^{\text{i}H_{xz1}\pi/2}R(\pi/2)$   \\
\label{TableLattice}
\end{tabular}
\end{center}
 where the coupling Hamiltonians are presented in Eq.~\ref{Eq_H3d} and Fig.~\ref{Fig_SupFig2}c. 
Extended forms of transition operators in the Rydberg coined DTQW would be: 
\begin{eqnarray}
\label{Eq_T3DCoined}
&&R_{\theta}=\text{e}^{\text{i} \theta \sigma_x}=\cos(\theta) \mathbbm{1}_{\{e,o\}}+\text{i} \sin(\theta) (\ket{e} \bra{ o}+\ket{o} \bra{ e})\\  \nonumber
&&T_{x}=  \sum\limits_{m_x}(\ket{m_x-1,e} \bra{ m_x,e}+\ket{m_x+1,o} \bra{ m_x,o})\otimes \mathbbm{1}_{y,z}\\ \nonumber 
&&T_{y}= \sum\limits_{m_y}   (\ket{m_y-1,e} \bra{ m_y,e }+  \ket{m_y+1,o} \bra{ m_y,o })\otimes \mathbbm{1}_{x,z} \\ \nonumber
&&T_{z}=   \sum\limits_{m_z}  (\ket{m_z-1,e} \bra{ m_z,e }+  \ket{m_z+1,o} \bra{ m_z,o })\otimes\mathbbm{1}_{x,y}\\ \nonumber
&&T_{xy}=\sum\limits_{m_x,m_y} (  \ket{m_x-1,m_y+1,e} \bra{ m_x,m_y,e }  + \ket{m_x+1,m_y-1,o} \bra{ m_x,m_y,o })\otimes\mathbbm{1}_{z}\\ \nonumber
&&T_{xz}=\sum\limits_{m_x,m_z} (  \ket{m_x-1,m_z-1,e} \bra{ m_x,m_z,e } + \ket{m_x+1,m_z+1,o} \bra{ m_x,m_z,o })\otimes\mathbbm{1}_{y}\\ \nonumber
&&T_{xyz}=\sum\limits_{{\bf m}} (  \ket{m_x-1,m_y-1,m_z-1,e} \bra{ m_x,m_y,m_z,e }  + \ket{m_x+1,m_y+1,m_z+1,o} \bra{ m_x,m_y,m_z,o }),
\end{eqnarray}  
where $\sigma_x$ is the Pauli operator in the odd-even basis.

\begin{figure}
\centering
\scalebox{0.50}{\includegraphics*{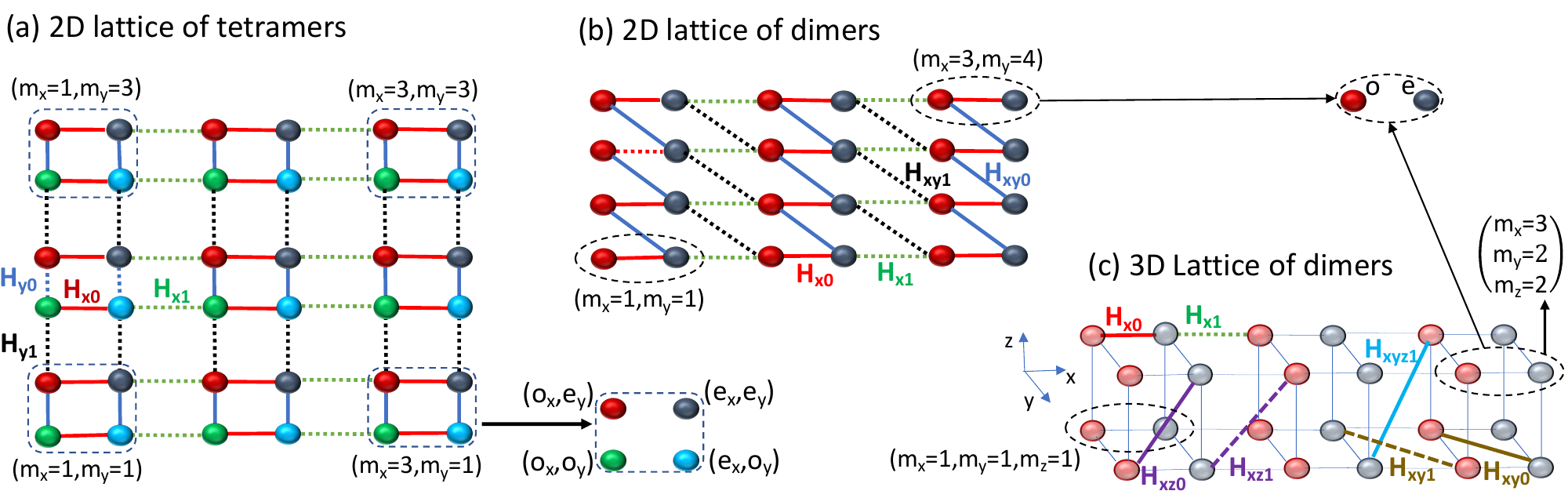}}  
\caption{Multi-dimensional DTQW. (a) Kronecker  multiplication of 1D QW leads to a 2D lattice of tetramers and could trivially be extended to a 3D lattice of octamers.
 Extension to the (b) 2D and (c) 3D lattice of dimers provides a non-separable multi-dimensional Rydberg DTQW. The dimers are labeled by $(m_x,m_y,m_z)$ while the sub-lattice elements are named by odd $o$ and even $e$ indices.}\label{Fig_SupFig2}  
\end{figure}

\end{widetext}

 \section*{A2: Topologically protected edge-state and Floquet topological insulators }
 \label{Sec_TopIns}
Application of SSH models in making  topological matters are vastly studied. This section discusses the implementation of the Floquet topological insulators (FTI) with the Rydberg coinless DTQW coupling Hamiltonians of Eq.~\ref{Eq_2DTetramers}, \ref{Eq_H3d} and re-drive the coined DTQW FTI \cite{Kit10} with the modified Rydberg coin operator of Eq.~\ref{Eq_T3DCoined}.
The robustness of the topological insulator/edge-state schemes against fluctuations of hopping angles presented in Fig. \ref{Fig_TopolEnergy}, \ref{Fig_Topol2DEdgeCoined}, \ref{Fig_Topol3DEdgeCoined}, allows fast operations with lower fidelity requirement, see Fig. \ref{Fig_Confignment}, \ref{Fig_Fidelity}. This would significantly enhance the number of coherent QW steps over larger lattices making the Rydberg proposal an ideal platform for large-scale  simulation of multi-dimensional topological insulators.

 \subsection*{Fourier transformed Hamiltonian} 
 For the QWs on an infinitely extended lattice, or a lattice with periodic torus boundary conditions, the QW operators can be transformed into quasi-momentum space $\tilde{H}$. The Fourier transformation of the odd $o$ and even $e$ sites  in 1D is given by
   \begin{eqnarray}
   \label{Eq_FourierElements}
&&\ket{m,e} =\frac{1}{\sqrt{\frac{N}{2}}}\sum \limits_{k} \text{e}^{\text{i}km} \ket{k,e}  \\ \nonumber
&&\ket{m,o} =\frac{1}{\sqrt{\frac{N}{2}}} \sum \limits_{k} \text{e}^{\text{i}km-\text{i}k\bar{a}_{1}}  \ket{k,o} \\ \nonumber
\end{eqnarray}
 where $ \ket{{k},e}= \ket{k}\otimes  \ket{e}$. Also  $\bar{a}_{1}=\frac{a_{1}}{a_{0}+a_{1}}$ with $a_{0}$, $a_{1}$ being the inter- and intra-dimer lattice constants.
 The wave-number is chosen to take on values from the first Brillouin zone $k=l\frac{2\pi}{N/2}$ with $1\leq l \leq {N}/{2}$. 
The Fourier transformed Hamiltonian would be obtained by replacing Eq.~\ref{Eq_FourierElements} into the QW Hamiltonians of Eq.~\ref{Eq_H}, simplified to 
  \begin{eqnarray}
&& \tilde{H}_0=\sum_{k}\begin{pmatrix}
0 &  \text{e}^{\text{i}{k}\bar{a}_{1}} \\
 \text{e}^{-\text{i}{k}\bar{a}_{1}} & 0  
\end{pmatrix}\ket{k}\bra{k}\\ \nonumber
&&\tilde{H}_1=\sum_{k}\begin{pmatrix}
0 &  \text{e}^{-\text{i}k(1-\bar{a}_{1})} \\
 \text{e}^{\text{i}k(1-\bar{a}_{1})} & 0  
\end{pmatrix}\ket{k}\bra{k},
 \end{eqnarray}
 with matrices being presented in the $\{\ket{o},\ket{e}\}$ basis.

\subsection*{Edge states in 1D}
\begin{figure}
\centering
\scalebox{0.38}{\includegraphics*[viewport=52 230 700 530]{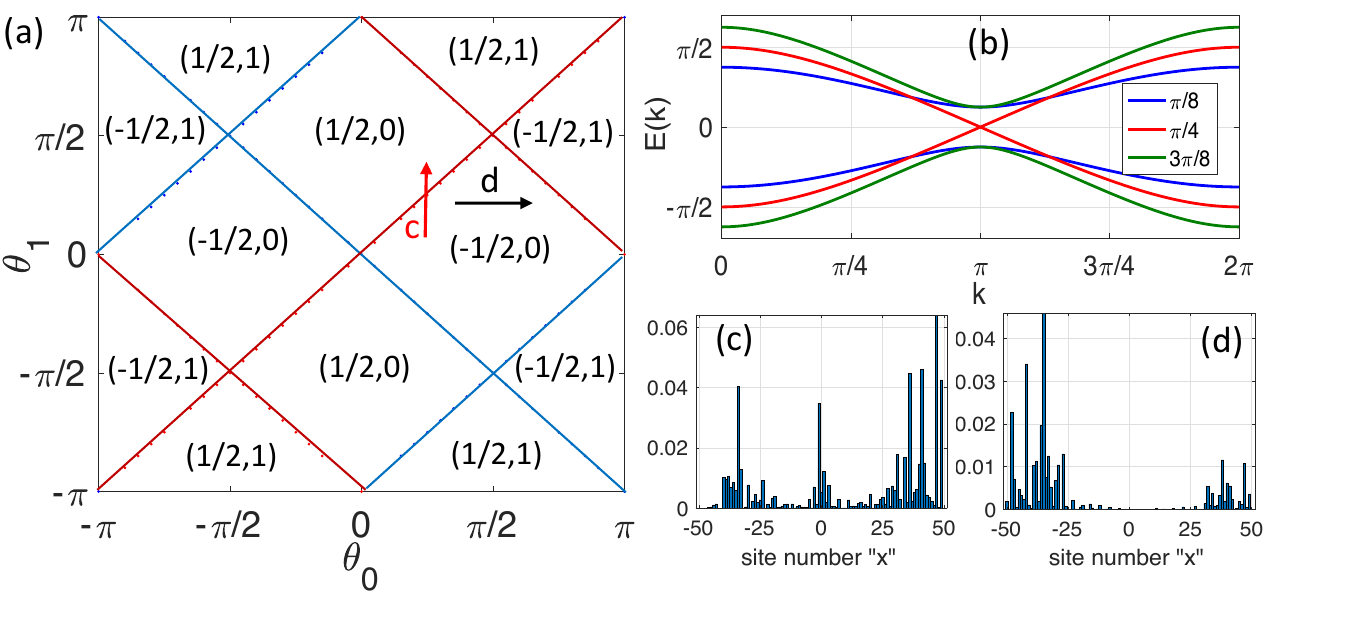}} 
\caption{ Topological edge state in 1D coinless DTQW.
(a) Topological invariants (winding number $\nu_0,\nu_{\pi}$) associated with the hopping angles  $\theta_0$ and  $\theta_1$, see Eq.~\ref{Eq_Winding}. 
(b) Band structure of the quantum walk for $\theta_0 = \pi/4$, $\theta_1=[\frac{\pi}{8},\frac{\pi}{4},\frac{3\pi}{8}]$. 
(c,d) shows the walker's wave-function after 100 steps for the space-dependent hopping angles explained in Eq.~\ref{Eq_angle} and over the range depicted by arrows in (a). The walker is initialized at the centre $x=0$ where the  sharp change of $\theta$  happens $w=0.1a_0$. In (c) the positive and negative sites have different winding numbers, providing the topological phase transition and leading to bound state at the centre. In (d) the evolution is adiabatic (no crossing of the  phase boundary) and hence  no population traps at the centre. 
  }\label{Fig_TopolEnergy}
\end{figure}

 The Floquet topological phases of Rydberg coinless discrete-time quantum walk can be accessed by looking at the full-time evolution of the walk.  
 The  quantum operator  in momentum basis
 \begin{eqnarray}
 \tilde{W}_{\text{eff}}=\text{e}^{\text{i}\frac{\theta_0}{2} \tilde{H}_0}\text{e}^{\text{i}\theta_1 \tilde{H}_1} \text{e}^{\text{i} \frac{\theta_0}{2} \tilde{H}_0}
 \label{Eq_U}
 \end{eqnarray}
   can be written as $\tilde{W}_{\text{eff}}=\text{e}^{\text{i}\tilde{H}_{\text{eff}}T}$, where $T$ is the period of applying a set of QW operators of Eq.~\ref{Eq_U}.
 In a lattice of dimers, $\tilde{H}_{\text{eff}}$ has two bands and hence  the effective Hamiltonian could be written as    
 \begin{equation}
 \tilde{H}_{\text{eff}}=\sum \limits_{k}  E(k)  \bm{n}(k).\pmb{\sigma} \, \ket{k} \bra{ k}
 \end{equation}
 where $\pmb{\sigma}=(\sigma_x,\sigma_y,\sigma_z)$ is the vector of Pauli matrices operating in the odd-even basis $\{  \ket{o}, \ket{e} \}$ and the $\bm{n}(k)=(n_x,n_y,n_z)$ defines the quantization axis for the spinor eigenstates at each momentum $k$.   The quantization axis of the eigen states are given by\\
$n_x=-\frac{(\sin \theta_0 \cos \theta_1  \cos(k\bar{a}_1)+\cos (k+k\bar{a}_1) \cos \theta_0 \sin \theta_1)}{\sin E(k)}$, $n_y=$\\
$\frac{ - \cos \theta_0 \sin \theta_1 \cos k \sin(k\bar{a}_1) +\sin \theta_0 \sin k \cos(k\bar{a}_1) - \sin \theta_0 \cos \theta_1 \sin(k\bar{a}_1)}{\sin E(k)}$, 
$n_z=0$.\\
 The quasi-energy band structure is formed by discrete translational invariance in time, which is given by
 \begin{equation}
 \cos E(k)= \cos \theta_0 \cos \theta_1-\sin \theta_0 \sin \theta_1 \cos(k),
 \end{equation}
and plotted in Fig.~\ref{Fig_TopolEnergy}b. 

In 1D, the topological phase is originated from the {\it chiral symmetry}. 
The chiral symmetry exists, if there is an antiunitary operator $\Gamma$ fulfilling the following relation $\Gamma H_{\text{eff}}\Gamma^{\dagger}=-H_{\text{eff}}$. While the trivial QW operator $\tilde{W}_{\text{eff}}=\text{e}^{\text{i}\theta_1 \tilde{H}_1}\text{e}^{\text{i}\theta_0 \tilde{H}_0}$ does not show the chiral symmetry, splitting one  of the steps  and moving the time frame to either of the following forms
 \begin{eqnarray}
  \label{Eq_WTopol}
 \tilde{W}_{\text{eff}}=\text{e}^{\text{i}\frac{\theta_0}{2} \tilde{H}_0}\text{e}^{\text{i}\theta_1 \tilde{H}_1} \text{e}^{\text{i} \frac{\theta_0}{2} \tilde{H}_0}\\ \nonumber
 \tilde{W}'_{\text{eff}}=\text{e}^{\text{i}\frac{\theta_1}{2} \tilde{H}_1}\text{e}^{\text{i}\theta_0 \tilde{H}_0} \text{e}^{\text{i} \frac{\theta_1}{2} \tilde{H}_1},
 \end{eqnarray}
the quantum walk do exhibit chiral symmetry with $\Gamma=\sigma_z$. 
 
The chiral symmetry forces the $\bm{n}(\theta_0,\theta_1,k)$ to rotate in the plain perpendicular to $\mathbf{\Gamma}$. The corresponding topological invariant is the number of times $\bm{n}(\theta_0,\theta_1,k)$ winds around the origin, as the quasi-momentum $k$ runs in the first Brillouin zone, called the {\it winding number}
 \begin{equation}
 \nu=\frac{1}{2\pi}\int \limits_{0}^{2\pi} \frac{1}{|\pmb{n}|^2} (n_x \partial_k  n_y -n_y \partial_k  n_x )\text{d}k.
 \label{Eq_Winding}
 \end{equation}
The two nonequivalent shifted time-frames $\tilde{W}_{\text{eff}}$, $\tilde{W}'_{\text{eff}}$ (Eq.~\ref{Eq_WTopol}),  lead to two winding numbers $\nu, \nu'$.  The two invariants $\nu_0=(\nu+\nu')/2$, $\nu_{\pi}=(\nu-\nu')/2$ would completely describe the topology. The phase diagram of the winding number  is plotted in Fig.~\ref{Fig_TopolEnergy}a.
 The band structure in Fig.~\ref{Fig_TopolEnergy}b are made of the energy eigenvalues $E(k,\theta_0,\theta_1)$. Manipulating the hopping angles  $\theta_0,\theta_1$ within the distinguished regions of Fig.~\ref{Fig_TopolEnergy}a, would force the system to  continuously transit between the band structures without
closing the energy gap, i.e. without changing the topological character of the system that is the winding number in here.  
At the borders that are separating distinct topological regions,  the band structure closes, see Fig.~\ref{Fig_TopolEnergy}b.

The topological character can be revealed at the boundary between the topologically distinct phases. 
To implement such a boundary one can apply inhomogeneous spatial hopping angle of the form 
\begin{equation}
\theta_i(x)=\frac{\theta_{i-}+\theta_{i+}}{2}+\frac{\theta_{i+}-\theta_{i-}}{2}\tanh(x/w)
\label{Eq_angle}
\end{equation}
where $w$ determines the spatial width of the phase transition region which defines the width of  the bound state. 
The variation of the hopping angle in the Rydberg system could be realized by different sets of exciting lasers or by applying space-dependent Stark-shift using an external field. The variation of the hopping angle $\theta$ shown by the red arrow in Fig.~\ref{Fig_TopolEnergy}a requires  4V/m electric field for an $\Omega/2\pi=2$MHz laser exciting $100S$ Rydberg state.
Fig.~\ref{Fig_TopolEnergy}c(d) shows the walker's wave-function after 100 steps for the cases with (without) the phase transition. The walker is initialized at $x=0$ in both cases. In Fig.~\ref{Fig_TopolEnergy}c, the positive and negative sites have different winding numbers. This would lead to the topological phase transition and hence form bound state at the centre. In Fig.~\ref{Fig_TopolEnergy}d, the evolution is adiabatic (i.e. there is no crossing of the  phase boundary) and no population traps at the centre.

 \subsection*{Floquet Topological Insulators with multi-dimensional DTQW}
 
  Different classes of Floquet topological insulators could be realized by the multi-dimensional coupling tessellations introduced in Sec.~\ref{Sec_multiD_QW}.
 The $nP$ excitation propagates unidirectionally and without backscattering along the edge, that is a line (surface) in 2D (3D) lattice, and eventually distributes uniformly along the boundary. Unlike in 1D, symmetry is not required for the presence of the topological phase in multi-dimensions.

 \subsubsection*{2D Topological insulator}

\begin{figure}
\centering
\scalebox{0.5}{\includegraphics*[viewport=1 110 810 530]{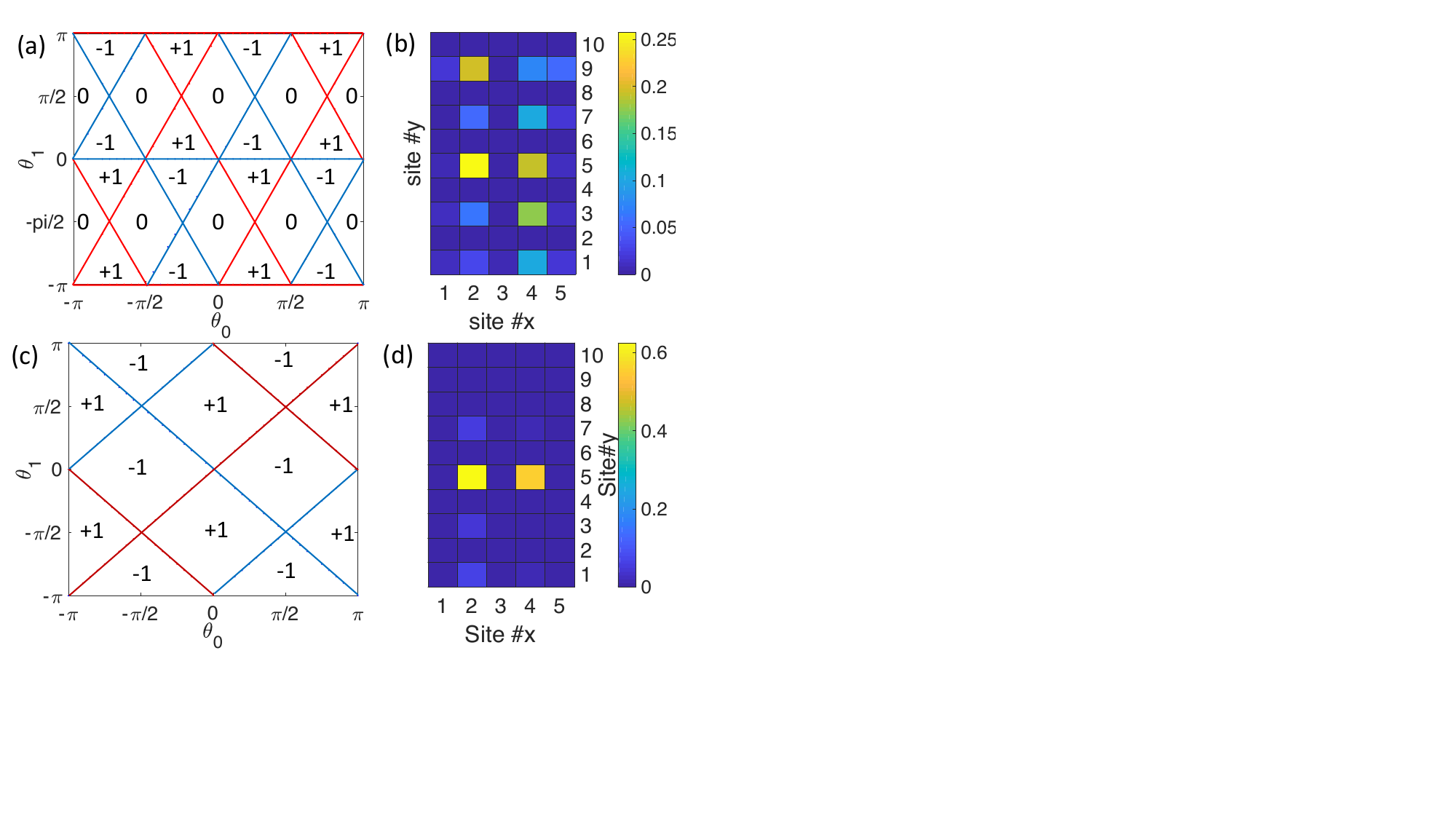}}
\caption{  Edge state in a 2D array of coined Rydberg DTQW.  TP edge states formed under coined quantum walk with the operator  of (a,b) $W_{\text{eff}}=T_{x}R_{\theta_0} T_{y} R_{\theta_1} T_{xy} R_{\theta_0}$ and (c,d) $W_{\text{eff}}=T_{y}R_{\theta_1} T_{x}  R_{\theta_0}$ with elements presented in Eq.~\ref{Eq_T3DCoined}. The topological phase diagram are plotted with the topological invariant that is the  (a) chern number and (b)  Rudner winding number. The energy gap of the two bands closes at 0  ($\pi$) at the red (blue) borders.  (b,d) Edge state is around the linear boundary at $x=2$ and $x=4$. It would distribute along the edge after large step numbers ($N_{step}=200$ in here). The  hopping angles are $(\theta_{0},\theta_{1})=(\pi/10,4\pi/10)$ inside $2\leq x\leq4$ and $(\theta_{0},\theta_{1})=(4\pi/10,\pi/10)$ outside the stripe $x<2$ and $x>4$. Periodic boundary condition is imposed along both $x$ and $y$ dimensions. Each site represents the sum of dimer's elements population.   }\label{Fig_Topol2DEdgeCoined}
\end{figure}

The {\it coined Rydberg DTQW} operators of Eq.~\ref{Eq_T3DCoined}, could ideally implement topological insulators with non-vanishing Chern numbers \cite{Kit10}. The concatenated operators $W_{\text{eff}}=T_{x}R_{\theta_0} T_{y} R_{\theta_1} T_{xy} R_{\theta_0}$ could be used for making topological insulators. To quantify the topological properties of this QW, the effective Hamiltonian in the momentum space is considered $\tilde{W}_{\text{eff}}=\text{e}^{\text{i}\tilde{H}_{\text{eff}}T}$. Like in the one-dimensional case discussed above, the discrete-time quantum walk is a stroboscopic simulator of the evolution generated by $\tilde{H}_{\text{eff}}$ at discrete-times.  The effective Hamiltonian would be 
 \begin{equation}
 \tilde{H}_{\text{eff}}=\sum \limits_{{\bm k}}  E({\bm k})  \bm{n}({\bm k}).\pmb{\sigma} \ket{{\bm k}} \bra{ {\bm k} }
 \end{equation}
 where $\pmb{\sigma}$, is the vector of Pauli matrices dealing with odd and even bases and the $\bm{n}({\bm k})$ defines the quantization axis for the spinor eigenstates at each momentum ${\bm k}$.  The topological invariant in this two dimensional Brillouin zone is given by the Chern number as 
  \begin{equation}
 C=\frac{1}{4\pi} \int \text{d}{\bm k}  \, {\bm n}.(\partial_{k_x} {\bm n} \times \partial_{k_y} {\bm n}   ).
 \label{Eq_Chern}
 \end{equation}
 The phase diagram of the quantum walk is plotted in Fig.~\ref{Fig_Topol2DEdgeCoined}a. 
Red and blue borders are associated with points where the energy gap closes at $E=0$ and $E=\pi$ respectively.

Like in the  1D case, TP edge states exist in 2D in spatial borders between  two regions with distinct topological phases. These states are propagating uni-directionally  in the bulk gaps and connect the bulk bands. A particle that is prepared in the superposition of the TP edge states, propagates coherently along the spatial border. The chirality of the edge states is topologically protected. In another words, their directions of propagation do not change  under continuous variation of the parameters of the system as long as the bulk gaps remain open. 
Fig.~\ref{Fig_Topol2DEdgeCoined}b,d  demonstrate TP propagating edge modes by considering an inhomogeneous 2D coined QW with spatial dependent coin angles.   Flat borders in the form of a strip geometry are considered in the 2D lattice, where the pair of coin angles inside and outside the strip belong to different topological phases. The  hopping angles are $(\theta_{0},\theta_{1})=(\pi/10,4\pi/10)$ inside $2\leq x\leq4$ and $(\theta_{0},\theta_{1})=(4\pi/10,\pi/10)$ outside the stripe $x<2$ and $x>4$.  The excitation is initialized on the borders as $( \ket{x=2,y=5,e}+ \ket{x=4,y=5,e})/\sqrt{2}$. The excitation would distribute along the border after large step numbers ( $N_{step}=200$ in here).

Unlike the time-independent QW in the time-dependent approach, topologically protected (TP) edge states could be formed even in the cases where the Chern number is zero for all the bands.
Fig.~\ref{Fig_Topol2DEdgeCoined}c,d discusses a simpl model of QW steps of $W_{\text{eff}}=T_{y}R_{\theta_1} T_{x}  R_{\theta_0}$ with operators being presented in Eq.~\ref{Eq_T3DCoined}.  In the  phase diagram of Fig.~\ref{Fig_Topol2DEdgeCoined}c, the Chern numbers of both bands are zero and the presented topological invariant is the Rudner winding number \cite{rud13}.

\begin{figure}
\centering
\scalebox{0.45}{\includegraphics*[viewport=10 130 1300 550]{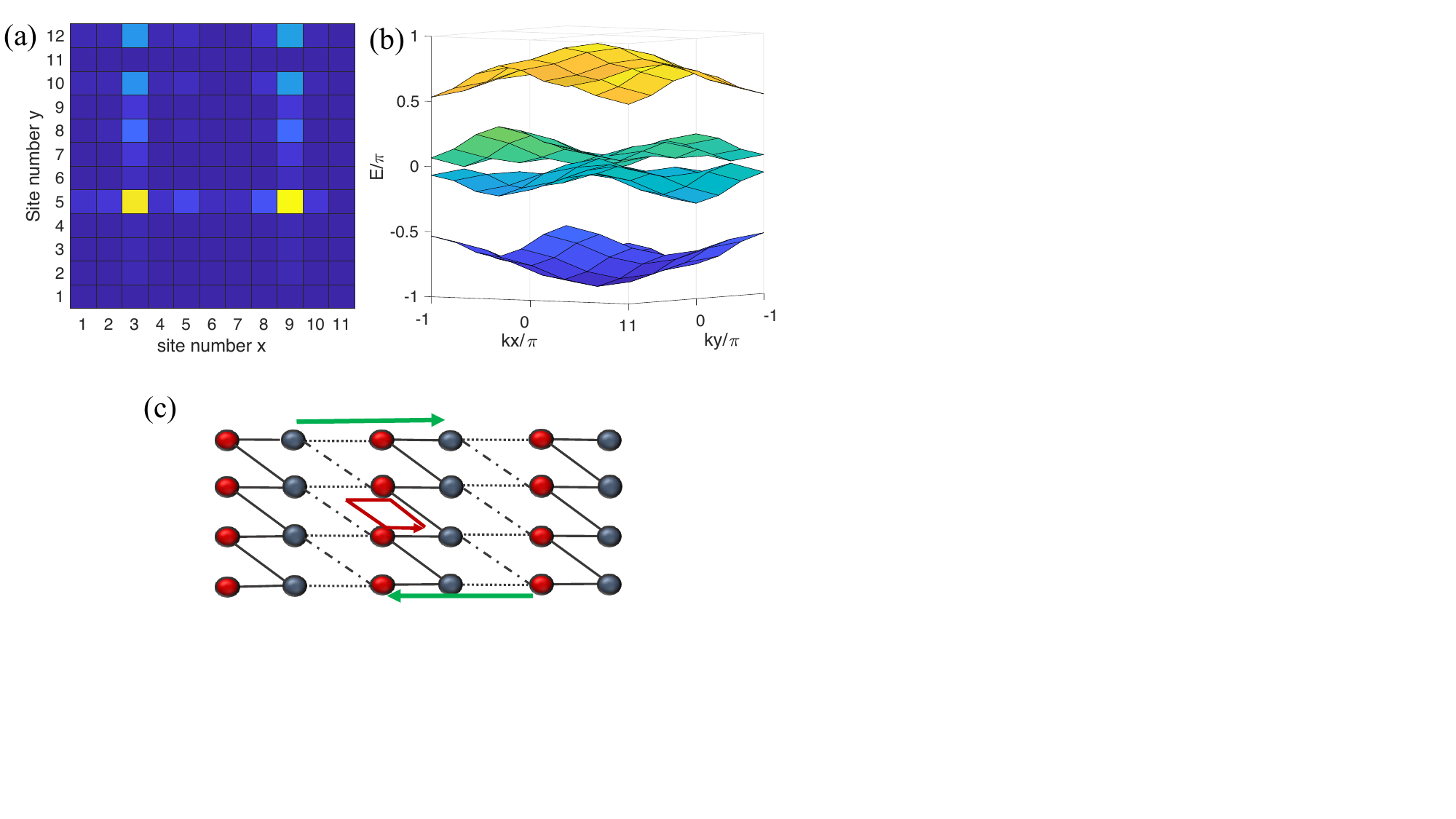}}
\caption{  Topological insulators  in a 2D array of (a,b) tetramers and (c) dimers with coinless DTQW.   
(a) Applying an inhomogeneous hopping angle along the $\hat{x}$ dimension with $(\theta_{x0},\theta_{x1})=(\pi/10,4\pi/10)$ inside $3< x\leq9$ and $(\theta_{x0},\theta_{x1})=(4\pi/10,\pi/10)$ outside the stripe ($x\le3$ and $x>9$), the excitation propagates unidirectionally along the borders of the stripe. The hopping angles  along the $y$ direction $(\theta_{y0},\theta_{y1})=(\pi/3,\pi/10)$ is not changing over different spatial regions. The population would distribute along the edge after large step numbers ( $N_{step}=209$ in here). 
 (b) The corresponding 4 bands associated with the tetramer sublattice are plotted for the hopping angle parameter choice inside the stripe. (c) In the 2D lattice of dimers all bulk states are localized, see the red arrow. A particle initially localized at any
site in the bulk, returns to its original position in one set of QW operations.  }\label{Fig_Topol2DEdge}
\end{figure}

The {\it coinless} models of a 2D lattice of tetramers as shown in Fig.~\ref{Fig_MultiD}a, could also be used for making
TP edge states. Here, the QW steps are effectively separable in two-dimensions. 
Fig.~\ref{Fig_Topol2DEdge}a shows that by applying inhomogeneous hopping angles  belonging to different regions of Fig.~\ref{Fig_TopolEnergy}a, along $\hat{x}$ dimension and using open(periodic) boundaries along $\hat{x}$($\hat{y}$) direction, one can form excitation currents solely on the borders.

Coin-less DTQW in a 2D lattice of dimers shown in Fig.~\ref{Fig_SupFig2}b could be used for the realization of the anomalous topological edge states \cite{rud13}. Here, the QW operators in different dimensions are not separable.
In an intuitive discussion, applying the operator $W=\text{e}^{\text{i}H_{x0}\pi/2} \text{e}^{\text{i} H_{xy1}\pi/2} \text{e}^{\text{i} H_{x1}\pi/2} \text{e}^{\text{i}H_{xy0}\pi/2}$  (with the Hamiltonians defined in Eq.~\ref{Eq_H3d}),  on the excitation initialized  on the border, leads to the clockwise transportation of the walker along the boarder as depicted by green arrows in Fig.~\ref{Fig_Topol2DEdge}c. This is while the initialized excitation in the bulk would go under unitary operator with no excitation transport, see red arrow in Fig.~\ref{Fig_Topol2DEdge}c. The exclusive conductance on the boundary provides the desired topological insulator. One can use inhomogeneous spatial angles with different Rudner winding numbers \cite{rud13} to design the shape of the edge state.

 \subsubsection*{3D Topological insulator}

\begin{figure}
\centering
\scalebox{0.35}{\includegraphics*[viewport=3 80 800 350]{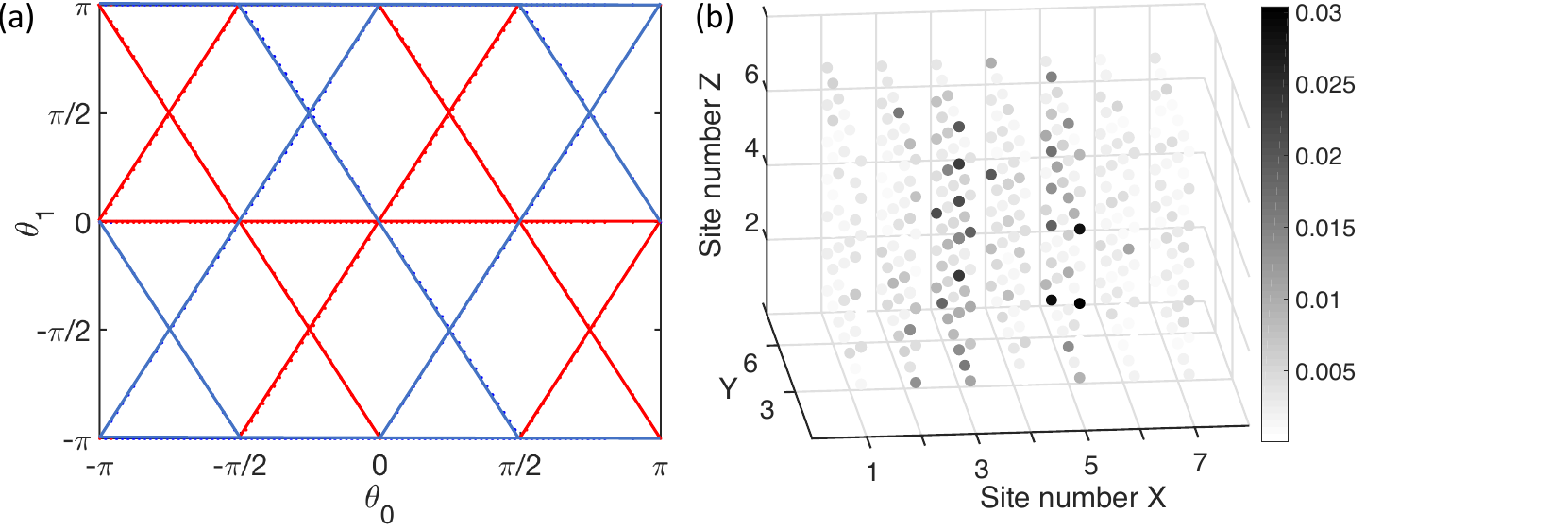}}
\caption{  Edge state in a 3D array of a coined Rydberg DTQW   $W_{\text{eff}}=T_{x}R_{\theta_0} T_{y} R_{\theta_1} T_{z} R_{\theta_0}$  with elements presented in Eq.~\ref{Eq_T3DCoined}. (a) The topological phase diagram are plotted where the two bands closes at $E=0$  ($E=\pi$) at the red (blue) borders.  (b) Edge state are formed on the surface boundaries $x=3$ and $x=5$. It would distribute on the edge surface after large step numbers ($N_{step}=200$ in here). The  hopping angles are $(\theta_{0},\theta_{1})=(4\pi/10,\pi/10)$ inside $3\leq x\leq5$ and $(\theta_{0},\theta_{1})=(\pi/10,4\pi/10)$ outside the stripe $x<3$ and $x>5$. Periodic boundary condition is imposed along $x$, $y$ and $z$ dimensions. Each site represents the sum of dimer's elements population.   }\label{Fig_Topol3DEdgeCoined}
\end{figure}

A  {\it 3D topological insulator} could be realized by the current proposal under a simple set of operators $W_{\text{eff}}=T_{x}  R_{\theta_0} T_{y}R_{\theta_1} T_{z}  R_{\theta_0}$, where the elements are defined in  Eq.~\ref{Eq_T3DCoined}.
The topological phase diagram is plotted in Fig.~\ref{Fig_Topol3DEdgeCoined}a, where the gap between the two bands closes at $E=0$  ($E=\pi$) at the red (blue) borders.  TP edge states in 3D exist along the surfaces  in spatial borders between two regions with distinct topological phases.  A particle, that is prepared in the superposition of the TP edge states, propagates coherently along the spatial border. 
Fig.~\ref{Fig_Topol3DEdgeCoined}b demonstrate TP propagating edge modes by considering an inhomogeneous 3D coined DTQW with spatially inhomogeneous coin angles. Here,  flat surface borders are considered. The pair of coin angles inside and outside the strip belongs to different topological phases. The  hopping angles are $(\theta_{0},\theta_{1})=(4\pi/10,\pi/10)$ inside $3\leq x\leq 5$ and $(\theta_{0},\theta_{1})=(\pi/10,4\pi/10)$ outside the stripe $x<3$ and $x>5$.  The excitation is initialized on the borders as $\ket{\psi_0}=( \ket{x=3,y=3,z=3,o}+ \ket{x=5,y=3,z=3,o})/\sqrt{2}$. The excitation would distribute over the border surface after large step numbers ( $N_{step}=200$ in here).

Topological insulators could also get implemented with 3D coinless DTQW model. For example $W=W_{x1}W_{xz1}W_{x0}W_{xz0}W_{xy0}W_{x0}W_{xy1}W_{x1}$ where $W_i=\exp(\text{i}H_i\theta_i)$ with Hamiltonians defined in Eq.~\ref{Eq_H3d} and uniform $\theta=\pi/2$ hopping angles on an open boundary lattice, result in an insulating bulk with exclusive excitation currents on the open boundaries.

\section*{A3: Self-interaction}

  Having one delocalized Rydberg excitation, the partial Rydberg population at different sites do not interact with each other. The nonlocal wave function only gives the probability of finding the single excitation at different sites.

Simulating the Schr\"odinger equation showed that partial population $P_{\ket{p}}$ at a specific site $i$ would induce the same population of auxiliary state $P_{\ket{s}}=P_{\ket{p}}$  in the  resonant sites $i$ and $j$.
Hence, when the {\it single} walker is not localized, the total population of the non-local auxiliary Rydberg state would add up to 1. 
Therefore, the absence of self interaction argument also applies to the induced single Rydberg auxiliary $\ket{s}$ state.

This argument is similar to the absence of self-interaction over a Rydberg polariton excited by a single photon in an atomic ensemble. 
The other point of similarity is in the Rydberg population dependence. In the Rydberg electromagnetically induced transparency (EIT), the ladder two-step excitation is derived by a faint quantum and a strong classical light. The population of Rydberg level is a function of the photon number in the quantum light. Similarly, in the proposed DTQW here, the maximum population of the auxiliary $\ket{s}$ state excited by the strong laser shining on  multiple ground state atoms is defined by the population of $\ket{p}$ state that makes the strong field in resonance with the laser transition. Therefore, the argument of the self-interaction also applies to the $\ket{s}$ single excitation.

 Following this argument, Eq.~\ref{Eq_HRy} only includes the inter-excitation interaction $V_{S-P}$ and does not consider self-interactions $V_{S-S}$ and $V_{P-P}$.

\begin{figure}
\centering
\scalebox{0.34}{\includegraphics*{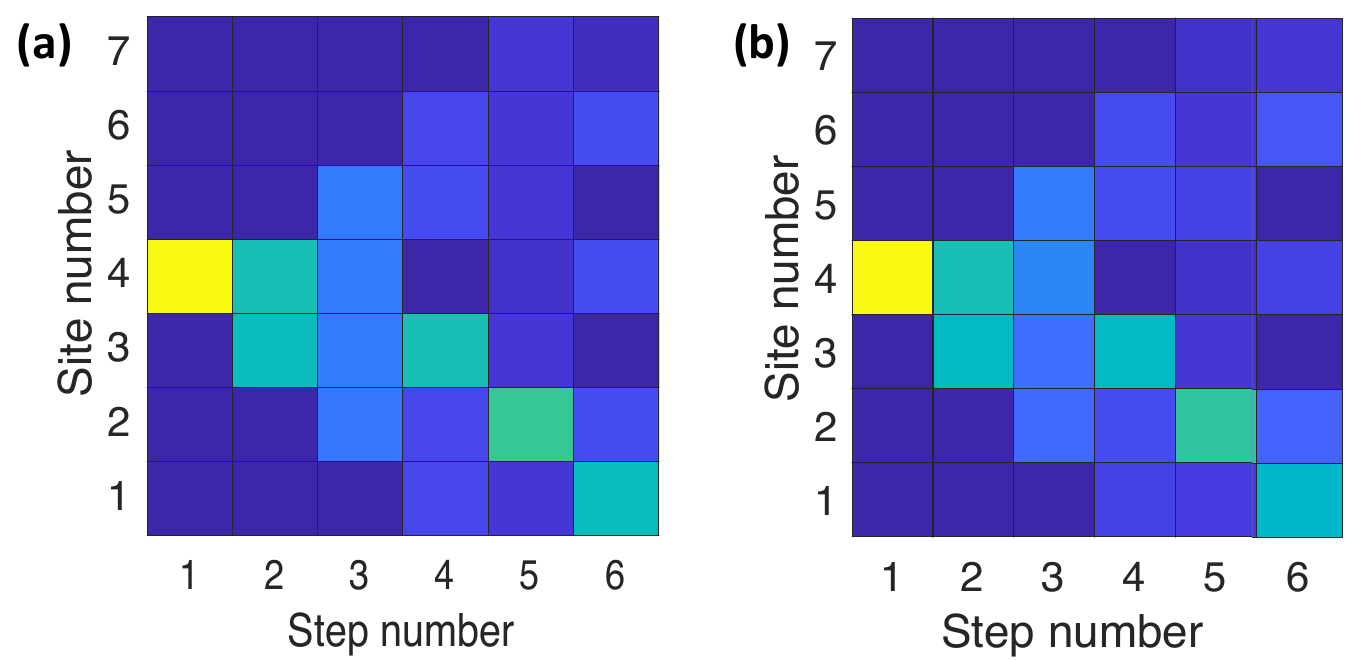}}
\caption{  Justifying the accuracy of Rydberg DTQW scheme. The histogram of a Hadamard ($\theta=\frac{\pi}{4}$) DTQW is plotted under the (a) analytical model of Eq.~\ref{Eq_W} and (b) the simulation of Schr\"odinger equation with the Rydberg Hamiltonian presented in Eq.~\ref{Eq_HRy}. The simulation in (b) is performed for the case exciting the $100S$ and $100P$ Rydberg states in a lattice with $(a_0,a_1)=(6,9)\mu$m with $\Omega=0.1V$ resulting in 98.5\% operation fidelity between (a) analytical and (b) numerical results.  }\label{Fig_Simulation}
\end{figure}


\begin{thebibliography}{100}


\bibitem{Aha93}
Y. Aharonov, L. Davidovich, and N. Zagury.
\newblock {\it Quantum random walks.}
\href{https://doi.org/10.1103/PhysRevA.48.1687} { Phys. Rev. A  {\bf 48}, 1687 (1993).}

\bibitem{Far98}
E. Farhi and S. Gutmann.
\newblock {\it Quantum computation and decision trees.}
\href{https://doi.org/10.1103/PhysRevA.58.915} { Phys. Rev. A {\bf 58}, 915 (1998).}

\bibitem{Kem03}
J. Kempe.
\newblock {\it Quantum random walks: an introductory overview.}
\href{ https://doi.org/10.1080/00107510902734722} { Contemp. Phys. {\bf 50}, 339 (2009).}

\bibitem{Dad18}
S. Dadras, A. Gresch, C.Groiseau, S. Wimberger, and G.~S
  Summy.
\newblock {\it Quantum walk in momentum space with a Bose-Einstein condensate.}
\href{https://doi.org/10.1103/PhysRevLett.121.070402} { Phys. Rev. Lett. {\bf 121}, 070402 (2018).}

\bibitem{Sum16}
G. Summy and S. Wimberger.
\newblock {\it Quantum random walk of a Bose-Einstein condensate in momentum space.}
\href{https://doi.org/10.1103/PhysRevA.93.023638} { Phys. Rev. A {\bf 93}, 023638 (2016).}

\bibitem{Pre15}
P.~M Preiss, et al., 
\newblock {\it Strongly correlated quantum walks in optical lattices.}
\href{https://doi.org/10.1126/science.1260364} { Science {\bf 347}, 1229 (2015).}

\bibitem{Por13}
R. Portugal.
\newblock {\em Quantum walks and search algorithms}.
\href{https://doi.org/10.1007/978-1-4614-6336-8}{ Springer, (2013).}

\bibitem{She03}
N. Shenvi, J. Kempe, and K~B. Whaley.
\newblock {\it Quantum random-walk search algorithm.}
\href{https://doi.org/10.1103/PhysRevA.67.052307} { Phys. Rev. A, {\bf 67}, 052307 (2003).}

\bibitem{Chi03}
A.~M Childs, R. Cleve, E. Deotto, E. Farhi, S. Gutmann, and
  D.~A Spielman.
\newblock {\it Exponential algorithmic speedup by a quantum walk.}
\href{https://doi.org/10.1145/780542.780552} { Proceedings of the thirty-fifth annual ACM symposium on
  Theory of computing, 59 (2003).}

\bibitem{Chi04}
A.~M Childs and J. Goldstone.
\newblock {\it Spatial search by quantum walk.}
\href{https://doi.org/10.1103/PhysRevA.70.022314} { Phys. Rev. A {\bf 70}, 022314 (2004).}

\bibitem{Por17}
R. Portugal and T.~D Fernandes.
\newblock {\it Quantum search on the two-dimensional lattice using the staggered
  model with hamiltonians.}
\href{https://doi.org/10.1103/PhysRevA.95.042341} {Phys. Rev. A {\bf 95}, 042341 (2017).}

\bibitem{Chi09}
A.~M Childs.
\newblock {\it Universal computation by quantum walk.}
\href{https://doi.org/10.1103/PhysRevLett.102.180501} { Phys. Rev. Lett. {\bf 102}, 180501 (2009).}

\bibitem{ken20}
V. Kendon, 
\newblock {\it How to compute using quantum walks.}
\href{https://doi.org/10.4204/EPTCS.315.1} {EPTCS {\bf 315}, 1 (2020)}.

\bibitem{Lov10}
N.~B Lovett, S. Cooper, M. Everitt, M. Trevers, and V. Kendon.
\newblock {\it Universal quantum computation using the discrete-time quantum walk.}
\href{https://doi.org/10.1103/PhysRevA.81.042330} { Phys. Rev. A {\bf 81}, 042330 (2010).}

\bibitem{Chi13}
A.~M Childs, D. Gosset, and Z. Webb.
\newblock {\it Universal computation by multiparticle quantum walk.}
\href{https://doi.org/10.1126/science.1229957} { Science {\bf 339}, 791 (2013).}

\bibitem{Sal12}
S.~El{\'\i}as V.-Andraca.
\newblock {\it Quantum walks: a comprehensive review.}
\href{https://doi.org/10.1007/s11128-012-0432-5} { Quantum Inf. Process {\bf 11}, 1015 (2012).}

\bibitem{Kit10}
T. Kitagawa, M.~S Rudner, E. Berg, and E. Demler.
\newblock {\it Exploring topological phases with quantum walks.}
\href{https://doi.org/10.1103/PhysRevA.82.033429} { Phys. Rev. A {\bf 82}, 033429 (2010).}

\bibitem{Sch09}
H.~Schmitz, et al.,
\newblock {\it Quantum walk of a trapped ion in phase space.}
\href{https://doi.org/10.1103/PhysRevLett.103.090504} { Phys. Rev. Lett. {\bf 103}, 090504 (2009).}

\bibitem{Zah10}
F.~Z\"ahringer, et al.,
\newblock {\it Realization of a quantum walk with one and two trapped ions.}
\href{https://doi.org/10.1103/PhysRevLett.104.100503} { Phys. Rev. Lett. {\bf 104}, 100503 (2010).}

\bibitem{Kar09}
M. Karski, et al.,
\newblock {\it Quantum walk in position space with single optically trapped atoms.}
\href{https://doi.org/10.1126/science.1174436} {Science {\bf 325}, 174 (2009).}

\bibitem{Wei11}
C. Weitenberg, et al.,
\newblock {\it Single-spin addressing in an atomic Mott insulator.}
\href{https://doi.org/10.1038/nature09827} { Nature {\bf 471}, 319 (2011).}

\bibitem{Fuk13}
T. Fukuhara, et al.,
\newblock {\it Microscopic observation of magnon bound states and their dynamics.}
\href{https://doi.org/10.1038/nature12541} { Nature {\bf 502}, 76 (2013).}

\bibitem{Man14}
Ji. Wang and K. Manouchehri.
\newblock {\em Physical implementation of quantum walks}.
\href{https://doi.org/10.1007/978-3-642-36014-5 }{ Springer (2013).}

\bibitem{Ber17}
H. Bernien, et al., 
\newblock {\it Probing many-body dynamics on a 51-atom quantum simulator.}
\href{https://doi.org/10.1038/nature24622} { Nature {\bf 551}, 579 (2017).}

\bibitem{Omr19}
A. Omran, et al., 
\newblock {\it Generation and manipulation of Schr{\"o}dinger cat states in Rydberg
  atom arrays.}
\href{https://doi.org/10.1126/science.aax9743} { Science {\bf 365}, 570 (2019).}

\bibitem{Zha11}
S~Zhang, F~Robicheaux, and M~Saffman.
\newblock {\it Magic-wavelength optical traps for Rydberg atoms.}
\href{https://doi.org/10.1103/PhysRevA.84.043408} { Phys. Rev. A {\bf 84}, 043408 (2011).}

\bibitem{Pio13}
MJ~Piotrowicz, et al., 
\newblock {\it Two-dimensional lattice of blue-detuned atom traps using a projected
  gaussian beam array.}
\href{https://doi.org/10.1103/PhysRevA.88.013420} { Phys. Rev. A {\bf 88}, 013420 (2013).}

\bibitem{Nog14}
F. Nogrette, et al., 
\newblock {\it Single-atom trapping in holographic 2D arrays of microtraps with
  arbitrary geometries.}
\href{https://doi.org/10.1103/PhysRevA.88.013420} { Phys. Rev. X {\bf 4}, 021034 (2014).}

\bibitem{Xia15}
T~Xia, et al., 
\newblock {\it Randomized benchmarking of single-qubit gates in a 2d array of
  neutral-atom qubits.}
\href{https://doi.org/10.1103/PhysRevLett.114.100503} { Phys. Rev. Lett. {\bf 114}, 100503 (2015).}

\bibitem{Zei16}
J. Zeiher, et al., 
\newblock {\it Many-body interferometry of a Rydberg-dressed spin lattice.}
\href{https://doi.org/10.1038/nphys3835} { Nature Physics {\bf 12}, 1095 (2016).}

\bibitem{Lie18}
V. Lienhard, et al., 
\newblock {\it Observing the space-and time-dependent growth of correlations in dynamically tuned synthetic ising models with antiferromagnetic interactions.}
\href{https://doi.org/10.1103/PhysRevX.8.021070} { Phys. Rev. X {\bf 8}, 021070 (2018).}

\bibitem{Nor18}
MA~Norcia, AW~Young, and AM~Kaufman.
\newblock {\it Microscopic control and detection of ultracold strontium in optical-tweezer arrays.}
\href{https://doi.org/10.1103/PhysRevX.8.041054} {Phys. Rev. X {\bf 8}, 041054 (2018).}

\bibitem{Yav09} D. D. Yavuz, N. A. Proite, and J. T. Green, {\it Nanometer-scale optical traps using atomic state localization},
 \href{https://doi.org/10.1103/PhysRevA.79.055401}{Phys. Rev. A {\bf 79}, 055401 (2009).}

\bibitem{Coo18}
A. Cooper, et al., 
\newblock {\it Alkaline-earth atoms in optical tweezers.}
\href{https://doi.org/10.1103/PhysRevX.8.041055} { Phys. Rev. X {\bf 8}, 041055 (2018).}

\bibitem{Hol19}
S. Hollerith, et al., 
\newblock {\it Quantum gas microscopy of Rydberg macrodimers.}
\href{https://doi.org/10.1103/PhysRevX.8.041055} { Science {\bf 364}, 664 (2019).}

\bibitem{Sas19}
S. Saskin, JT~Wilson, B. Grinkemeyer, and J.~D. Thompson.
\newblock {\it Narrow-line cooling and imaging of ytterbium atoms in an optical
  tweezer array.}
\href{https://doi.org/10.1103/PhysRevLett.122.143002} { Phys. Rev. Lett. {\bf 122}, 143002 (2019).}

\bibitem{Wan16}
Y. Wang, A. Kumar, T.-Y. Wu, and D.~S Weiss.
\newblock {\it Single-qubit gates based on targeted phase shifts in a 3D neutral
  atom array.}
\href{https://doi.org/10.1126/science.aaf2581} {Science {\bf 352}, 1562 (2016).}

\bibitem{Bar18}
D. Barredo, V. Lienhard, S. De~Leseleuc, T. Lahaye, and  A. Browaeys.
\newblock {\it Synthetic three-dimensional atomic structures assembled atom by atom.}
\href{https://doi.org/10.1038/s41586-018-0450-2} { Nature {\bf 561}, 79 (2018).}

\bibitem{Lev18}
H. Levine, et al., 
\newblock {\it High-fidelity control and entanglement of Rydberg-atom qubits.}
\href{https://doi.org/10.1103/PhysRevLett.121.123603} {Phys. Rev. Lett. {\bf 121}, 123603 (2018).}

\bibitem{Saf10}
M. Saffman, T.~G Walker, and K. M{\o}lmer.
\newblock {\it Quantum information with Rydberg atoms.}
\href{https://doi.org/10.1103/RevModPhys.82.2313} { Rev. Mod. Phys. {\bf 82}, 2313 (2010).}

\bibitem{Ada19}
CS~Adams, JD~Pritchard, and JP~Shaffer.
\newblock {\it Rydberg atom quantum technologies.}
\href{https://doi.org/10.1088/1361-6455/ab52ef} { J. Phys. B  {\bf 53}, 012002 (2019).}

\bibitem{Kha15}
M. Khazali, K. Heshami, and C. Simon.
\newblock {\it Photon-photon gate via the interaction between two collective Rydberg
  excitations.}
  \href{https://doi.org/10.1103/PhysRevA.91.030301}{{ Phys. Rev. A}, \textbf{91}, 030301 (2015).}

\bibitem{kha20}
M. Khazali and K. M{\o}lmer.
\newblock {\it Fast multiqubit gates by adiabatic evolution in interacting
  excited-state manifolds of Rydberg atoms and superconducting circuits.}
\href{https://doi.org/10.1103/PhysRevX.10.021054} {{ Phys. Rev. X}, \textbf{10}, 021054 (2020).}

\bibitem{kha21}M. Khazali, and W. Lechner. {\it "Electron cloud design for Rydberg multi-qubit gates."} \href{https://arxiv.org/abs/2111.01581}{ arXiv:2111.01581 (2021).}

\bibitem{khaz2020Rydberg}
M. Khazali.
\newblock {\it Rydberg noisy-dressing and applications in making soliton-molecules
  and droplet quasi-crystals.}
\href{https://doi.org/10.1103/PhysRevResearch.3.L032033} {{ Phys. Rev. Research} \textbf{3}, L032033 (2020)}.

\bibitem{khaz16}
Khazali, Mohammadsadegh. {\it Applications of Atomic Ensembles for Photonic Quantum Information Processing and Fundamental Tests of Quantum Physics.} \href{https://www.proquest.com/openview/b35882136d01bc6643040d8c66c410e6/1?cbl=18750&pq-origsite=gscholar}{Diss. University of Calgary (Canada), (2016).}


\bibitem{Kha21}
Khazali, M. {\it Quantum Information and Computation with Rydberg Atoms.} \href{https://doi.org/10.22051/IJAP.2021.34445.1188} {Iranian Journal of Applied Physics \textbf{10}, 19 (2021).}


\bibitem{Kha19}
M. Khazali, C.~R Murray, and T. Pohl.
\newblock {\em Polariton exchange interactions in multichannel optical networks.}
\href{https://doi.org/10.1103/PhysRevLett.123.113605} {{ Phys. Rev. Lett.}, \textbf{123}, 113605 (2019).}

\bibitem{Kha16}
M. Khazali, H.~W. Lau, A. Humeniuk, and C. Simon.
{\it Large energy superpositions via Rydberg dressing.}
\href{https://doi.org/10.1103/PhysRevA.94.023408} {{Phys. Rev. A}, \textbf{94}, 023408 (2016).}

\bibitem{Kha17}
M. Khazali, K. Heshami, and C. Simon.
\newblock Single-photon source based on Rydberg exciton blockade.
\href{https://doi.org/10.1088/1361-6455/aa8d7c} { J. Phys. B: At. Mol. Opt. Phys.
 {\bf 50}, 215301, (2017).}

\bibitem{Kha18}
M. Khazali.
\newblock Progress towards macroscopic spin and mechanical superposition via
  Rydberg interaction.
\href{https://doi.org/10.1103/PhysRevA.98.043836} { Phys. Rev. A {\bf 98}, 043836 (2018).}

\bibitem{Cot06}
R. C{\^o}t{\'e}, A. Russell, E.~E Eyler, and P.~L Gould.
\newblock {\it Quantum random walk with Rydberg atoms in an optical lattice.}
\href{https://doi.org/10.1088/1367-2630/8/8/156} { New J. Phys. {\bf 8}, 156 (2006).}

\bibitem{Les19} de Leseleuc, Sylvain, et al. {\it "Observation of a symmetry-protected topological phase of interacting bosons with Rydberg atoms."}
\href{https://doi.org/10.1126/science.aav9105} {Science {\bf 365}, 775 (2019).}


\bibitem{Bar15}
D. Barredo, et al., 
\newblock {\it Coherent excitation transfer in a spin chain of three Rydberg atoms.}
\href{https://doi.org/10.1103/PhysRevLett.114.113002} { Phys. Rev. Lett. {\bf 114}, 113002 (2015).}

\bibitem{Scho15}
DW~Sch{\"o}nleber, A. Eisfeld, M. Genkin, S~Whitlock, and S.  W{\"u}ster.
\newblock {\it Quantum simulation of energy transport with embedded Rydberg
  aggregates.}
\href{https://doi.org/10.1103/PhysRevLett.114.123005} {Phys. Rev. Lett. {\bf 114}, 123005 (2015).}

\bibitem{Ori18}
A~Pineiro Orioli, et al., 
\newblock {\it Relaxation of an isolated dipolar-interacting Rydberg quantum spin
  system.}
\href{https://doi.org/10.1103/PhysRevLett.120.063601} {Phys. Rev. Lett. {\bf 120}, 063601 (2018).}

\bibitem{Gun13}
G~G{\"u}nter, et al., 
\newblock {\it Observing the dynamics of dipole-mediated energy transport by
  interaction-enhanced imaging.}
\href{https://doi.org/10.1126/science.1244843} { Science {\bf 342}, 954 (2013).}

\bibitem{Sch15}
H~Schempp, et al., 
\newblock {\it Correlated exciton transport in Rydberg-dressed-atom spin chains.}
\href{https://doi.org/10.1103/PhysRevLett.115.093002} {Phys. Rev. Lett. {\bf 115}, 093002 (2015).}

\bibitem{Let18}
F. Letscher and . Petrosyan.
\newblock {\it Mobile bound states of Rydberg excitations in a lattice.}
\href{https://doi.org/10.1103/PhysRevA.97.043415} { Phys. Rev. A {\bf 97}, 043415 (2018).}

\bibitem{Wus11}
S~W{\"u}ster, C~Ates, A~Eisfeld, and JM~Rost.
\newblock {\it Excitation transport through Rydberg dressing.}
\href{https://doi.org/10.1088/1367-2630/13/7/073044} { New J. Phys. {\bf 13}, 073044 (2011).}

\bibitem{Dau16}
A. Dauphin, M. M{\"u}ller, and M.~A. Martin-Delgado.
\newblock {\it Quantum simulation of a topological Mott insulator with Rydberg atoms
  in a Lieb lattice.}
\href{https://doi.org/10.1103/PhysRevA.93.043611} { Phys. Rev. A {\bf 93}, 043611 (2016).}

\bibitem{And13}
Y. Ando.
\newblock {\it Topological insulator materials.}
\href{https://doi.org/10.7566/JPSJ.82.102001} { J. Phys. Soc. Jpn {\bf 82}, 102001 (2013).}

\bibitem{Cay13}
J. Cayssol, B. D{\'o}ra, F. Simon, and R.  Moessner.
\newblock {\it Floquet topological insulators.}
\href{ https://doi.org/10.1002/pssr.201206451} {pss (RRL)  {\bf 7}, 101 (2013).}

\bibitem{Kit09}
A. Kitaev.
\newblock {\it Periodic table for topological insulators and superconductors.}
\href{https://doi.org/10.1063/1.3149495} { AIP Conf Proc  {\bf 1134}, 22
  (2009).}

\bibitem{Pan20}
S~Panahiyan and S~Fritzsche.
\newblock {\it Toward simulation of topological phenomenas with one-, two-and
  three-dimensional quantum walks.}
\href{https://doi.org/10.1103/PhysRevA.103.012201} {Phys. Rev. A {\bf 103}, 012201 (2021).}

\bibitem{Rec13}
Mikael~C Rechtsman, et al., 
\newblock {\it Photonic floquet topological insulators.}
\href{https://doi.org/10.1038/nature12066} { Nature {\bf 496}, 196 (2013).}

\bibitem{Xia17}
L~Xiao, et al., 
\newblock {\it Observation of topological edge states in parity-time-symmetric
  quantum walks.}
\href{https://doi.org/10.1038/nphys4204} {Nature Phys. {\bf 13}, 1117 (2017).}

\bibitem{Muk17}
S. Mukherjee, et al., 
\newblock {\it Experimental observation of anomalous topological edge modes in a
  slowly driven photonic lattice.}
\href{https://doi.org/10.1038/ncomms13918} { Nature Com. {\bf 8}, 1 (2017).}



\bibitem{Amb15} A. Ambainis, R. Portugal, and N. Nahimov, {\it Spatial search on grids with minimum memory}, \href{https://doi.org/10.26421/QIC15.13-14-9}{Quantum Information and Computation {\bf 15}, 1233 (2015).}
\bibitem{Por15}  R. Portugal, S. Boettcher, and S. Falkner, {\it One-dimensional coinless quantum walks}, \href{https://doi.org/10.1103/PhysRevA.91.052319}{Phys. Rev. A {\bf 91}, 052319 (2015).}
\bibitem{Por16} R. Portugal, R. A. M. Santos, T. D. Fernandes and D. N. Goncalves, {\it The staggered quantum walk model}, \href{https://doi.org/10.1007/s11128-015-1149-z}{Quantum Information Processing {\bf 15}, 85 (2016).}
\bibitem{Port16} R. Portugal, {\it Staggered quantum walks on graphs}, \href{https://doi.org/10.1103/PhysRevA.93.062335}{Phys. Rev. A {\bf 93}, 062335 (2016).}
\bibitem{Portu17}R. Portugal, M. C. de Oliveira, and J. K. Moqadam, {\it Staggered quantum walks with Hamiltonians}, \href{https://doi.org/10.1103/PhysRevA.95.012328}{Phys. Rev. A {\bf 95}, 012328 (2017).}


\bibitem{su1979solitons}
W. P. Su, JR~Schrieffer, and Ao~J Heeger.
\newblock {\it Solitons in polyacetylene.}
\href{https://doi.org/10.1103/PhysRevLett.42.1698} { Phys. Rev. Lett. {\bf 42}, 1698 (1979).}


\bibitem{Mich17}T M Michelitsch et al.,{\it Recurrence of random walks with long-range steps generated by fractional Laplacian matrices on regular networks and simple cubic lattices},
\href{https://doi.org/10.1088/1751-8121/aa9008}{J. Phys. A: Math. Theor. {\bf 50}, 505004 (2017)}


\bibitem{Moq18} Moqadam, J. Khatibi, and Ali T. Rezakhani. 
{\it Boundary-induced coherence in the staggered quantum walk on different topologies.} 
\href{https://doi.org/10.1103/PhysRevA.98.012123} {Phys. Rev. A {\bf 98}, 012123 (2018).}

\bibitem{Xia10}D. Xiao, M.-C. Chang, and Q. Niu, {\it Berry phase effects on electronic properties}, \href{10.1103/RevModPhys.82.1959}{Rev. Mod. Phys. {\bf 82}, 1959 (2010).}

\bibitem{Gru14}F. Grusdt and M. Honing, {\it Realization of fractional Chern insulators in the thin-torus limit with ultracold bosons}, \href{https://doi.org/10.1103/PhysRevA.90.053623}{Phys. Rev. A {\bf 90}, 053623 (2014)}

\bibitem{Lac16}M. Lacki, et al., {\it Quantum Hall physics with cold atoms in cylindrical optical lattices},
\href{https://doi.org/10.1103/PhysRevA.93.013604}{Phys. Rev. A {\bf 93}, 013604 (2016).}

\bibitem{Schl07}
M.~A Schlosshauer.
\newblock {\em Decoherence: and the quantum-to-classical transition}.
\href{https://doi.org/10.1007/978-3-540-35775-9} {Springer  (2007).}

\bibitem{Alb14}Alberti, Andrea, et al. {\it "Decoherence models for discrete-time quantum walks and their application to neutral atom experiments."} \href{https://doi.org/10.1088/1367-2630/16/12/123052}{New Journal of Physics {\bf 16}, 123052 (2014).}


\bibitem{Ros19}
R.~B Hutson, Aet al., 
\newblock {\it Engineering quantum states of matter for atomic clocks in shallow
  optical lattices.}
\href{https://doi.org/10.1103/PhysRevLett.123.123401} { Phys. Rev. Lett. {\bf 123}, 123401 (2019).}

\bibitem{Kau12}
A.~M Kaufman, B.~J Lester, and C.~A Regal.
\newblock {\it Cooling a single atom in an optical tweezer to its quantum ground
  state.}
\href{https://doi.org/10.1103/PhysRevX.2.041014} { Phys. Rev. X {\bf 2}, 041014 (2012).}

\bibitem{Tho13}
J.~D. Thompson, et al.,
\newblock {\it Coherence and raman sideband cooling of a single atom in an optical
  tweezer.}
\href{https://doi.org/10.1103/PhysRevLett.110.133001} { Phys. Rev. Lett. {\bf 110}, 133001 (2013).}

\bibitem{Bel13}
N. Belmechri, et al., 
\newblock {\it Microwave control of atomic motional states in a spin-dependent
  optical lattice.}
\href{https://doi.org/10.1088/0953-4075/46/10/104006} { J.  Phys. B {\bf 46}, 104006 (2013).}

\bibitem{Wan19}Wang, Kunpeng, et al. {\it "Preparation of a heteronuclear two-atom system in the three-dimensional ground state in an optical tweezer."} 
\href{https://doi.org/10.1103/PhysRevA.100.063429}{Phys. Rev. A {\bf 100}, 063429 (2019).}


\bibitem{Bar20} D. Barredo, et al., 
{\it Three-Dimensional Trapping of Individual Rydberg Atoms in Ponderomotive Bottle Beam Traps,} 
\href{https://doi.org/10.1103/PhysRevLett.124.023201}{Phys. Rev. Lett. {\bf 124}, 023201 (2020).}

\bibitem{Gra19} T. M. Graham, et al., 
 {\it Rydberg Mediated Entanglement in a Two-Dimensional Neutral Atom Qubit Array,} 
\href{https://doi.org/10.1103/PhysRevLett.123.230501}{Phys. Rev. Lett. {\bf 123}, 230501 (2019).}


\bibitem{Wil19} Wilson, J., et al., 
{\it Trapped arrays of alkaline earth Rydberg atoms in optical tweezers. }\href{https://arxiv.org/abs/1912.08754}{ arXiv:1912.08754 (2019). }

\bibitem{Bet09}
II~Beterov, II~Ryabtsev, DB~Tretyakov, and VM~Entin.
\newblock {\it Quasiclassical calculations of blackbody-radiation-induced
  depopulation rates and effective lifetimes of Rydberg ns, np, and nd
  alkali-metal atoms with n<80.}
\href{https://doi.org/10.1103/PhysRevA.79.052504} { Phys. Rev. A {\bf 79}, 052504 (2009).}



 \bibitem{Sig17}Signoles, A., et al., 
 {\it Coherent transfer between low-angular-momentum and circular Rydberg states},
 \href{https://doi.org/10.1103/PhysRevLett.118.253603}{ Phys. Rev. Lett. {\bf 118}, 253603 (2017).}
 
 \bibitem{Car20}R. Cardman and G. Raithel, {\it Circularizing Rydberg atoms with time-dependent optical traps, }
 \href{https://doi.org/10.1103/PhysRevA.101.013434}{Phys. Rev. A {\bf 101}, 013434 (2020).}

\bibitem{Ngu18}
T.~Long Nguyen, et~al. 
\newblock {\it Towards quantum simulation with circular Rydberg atoms.}
\href{https://doi.org/10.1103/PhysRevX.8.011032} {Phys. Rev. X {\bf 8}, 011032, (2018).}

\bibitem{Kwo17}
M. Kwon, M.~F Ebert, T.~G Walker, and M~Saffman.
\newblock Parallel low-loss measurement of multiple atomic qubits.
\href{https://doi.org/10.1103/PhysRevLett.119.180504} { Phys. Rev. Lett. {\bf 119}, 180504 (2017).}




\bibitem{Cov19}
J.~P Covey, I.~S Madjarov, A. Cooper, and M. Endres.
{\it 2000-times repeated imaging of strontium atoms in clock-magic tweezer  arrays.}
\href{https://doi.org/10.1103/PhysRevLett.122.173201} {Phys. Rev. Lett. {\bf 122}, 173201 (2019).}



\bibitem{Wie16}
B~J Wieder and CL~Kane.
\newblock {\it Spin-orbit semimetals in the layer groups.}
\href{https://doi.org/10.1103/PhysRevB.94.155108} { Phys. Rev. B {\bf 94}, 155108 (2016).}



\bibitem{saj18}
M. Sajid, et al., 
\newblock {\it Creating Floquet Chern insulators with magnetic quantum walks.}
\href{https://doi.org/10.1103/PhysRevB.99.214303} {Phys. Rev. B {\bf 99} 214303  (2019).}


\bibitem{Nei03} S. Neil, J. Kempe, and K. Whaley. {\it "Quantum random-walk search algorithm."} 
\href{https://doi.org/10.1103/PhysRevA.67.052307}{Phys. Rev. A {\bf 67}, 052307 (2003).}

\bibitem{Seg00} Luc Segoufin, Victor Vianu,
{\it Querying Spatial Databases via Topological Invariants,}
\href{https://doi.org/10.1006/jcss.2000.1712.}{Journal of Computer and System Sciences {\bf 61}, 270 (2000).}

\bibitem{Cle02} Clementini, E. {\it "Topological relations in spatial databases"}, \href{https://doi.org/10.1201/9781420040814.Ch3d}{Intelligent Systems: Technology and Applications {\bf 4},  47 (2002).}


\bibitem{rud13}
M.~S Rudner, N.~H Lindner, E. Berg, and M. Levin.
\newblock{\it Anomalous edge states and the bulk-edge correspondence for  periodically driven two-dimensional systems.}
\href{https://doi.org/10.1103/PhysRevX.3.031005} {\it Phys. Rev. X {\bf 3}, 031005 (2013).}



\end{thebibliography}
\end{document}